\documentclass[aps,prd,preprint,superscriptaddress,amsmath,amssymb,showpacs]{revtex4-1}
\usepackage{dcolumn}
\usepackage{graphicx}
\usepackage{float}
\usepackage{physics}
\usepackage[colorlinks=true,allcolors=blue]{hyperref}
\usepackage{mathrsfs}
\usepackage{inputenc}

\begin{document}
\title{Thermodynamics and energy loss in D dimensions from holographic QCD model}

\author{Zhou-Run Zhu}
\email{zhuzhourun@mails.ccnu.edu.cn}
\affiliation{Institute of Particle Physics and Key Laboratory of Quark and Lepton Physics (MOS), Central China Normal University,
Wuhan 430079, China}

\author{Jun-Xia Chen}
\email{chenjunxia@mails.ccnu.edu.cn}
\affiliation{Institute of Particle Physics and Key Laboratory of Quark and Lepton Physics (MOS), Central China Normal University,
Wuhan 430079, China}

\author{Xian-Ming Liu}
\email{200931140009@mail.bnu.edu.cn}
\affiliation{Institute of Particle Physics and Key Laboratory of Quark and Lepton Physics (MOS), Central China Normal University, Wuhan 430079, China}
\affiliation{Department of Physics, Hubei Minzu University, Enshi 445000, China}

\author{Defu Hou }
\thanks{Corresponding author}
\email{houdf@mail.ccnu.edu.cn}
\affiliation{Institute of Particle Physics and Key Laboratory of Quark and Lepton Physics (MOS), Central China Normal University,
Wuhan 430079, China}

\date{\today}

\begin{abstract}
We consider the holographic QCD model with a planar horizon in the D dimensions with different consistent metric solutions. We investigate the black hole thermodynamics, phase diagram and equations of state (EoS) in different dimensions. The temperature and chemical potential dependence of the drag force and diffusion coefficient also have been studied. From the results, the energy loss of heavy quark shows an enhancement near the phase transition temperature in D dimensions. This finding illustrates that the energy loss of heavy quark has a nontrivial and non-monotonic dependence on temperature. Furthermore, we find the heavy quark may lose less energy in higher dimension. The diffusion coefficient is larger in higher dimension.
\end{abstract}

\maketitle

\section{Introduction}\label{sec:01_intro}
The extreme environment of high temperature and energy density created by heavy ion collision experiments at RHIC and LHC \cite{Arsene:2004fa,Adcox:2004mh,Back:2004je,Adams:2005dq,Aad:2013xma} may provide the best chance to characterize the strongly coupled plasma. The investigation of the phase structure of quantum chromodynamics (QCD) is significant and challenging. As known to all, QCD is in the confinement regime at low temperature/density, while in the deconfinement regime at high temperature/density. A phase transition may exist between confinement phase and deconfinement phase. The phase diagram of QCD displays a rich information \cite{Stephanov:2004wx,Ding:2015ona} which involves the phase transition, the location of the critical point, etc. Although lattice QCD is a powerful tool to understand the strongly interacting properties and provides reliable results at zero chemical potential, it does not work in finite chemical potential case. Furthermore, the perturbative QCD techniques become unreliable around the phase transition since coupling constant becomes very strong. It is generally believed that a strongly coupled quark-gluon plasma (QGP) produced in the experiments behaves as a nearly perfect liquid \cite{Gyulassy:2004zy}. The QGP liquid with a small $\eta/s$ (the shear viscosity to entropy density ratio) is close to the ideal relativistic hydrodynamic limit. The estimated values of $\eta/ s$ based on the hydrodynamic model and the experimental data are between $0.095$ and $0.2$ \cite{Luzum:2008cw,Ryu:2015vwa,Gale:2012rq}. This indicates that we need a powerful tool to study QCD. AdS/CFT correspondence may be an interesting tool. In holography, the value of $\eta/s$ is $1/4\pi$ \cite{Policastro:2001yc,Buchel:2003tz,Kovtun:2004de} which is consistent with experimental data.

In the limit of large 't Hoot coupling $\lambda \equiv g^{2} N_{c}$, AdS/CFT correspondence \cite{Witten:1998qj,Gubser:1998bc,Maldacena:1997re} can be described by the duality between gravity theory in the AdS spacetime and $\mathcal{N} = 4$ SYM theory with a gauge group on the AdS boundary. In holography, an external quark is dual to a string dangling from the boundary towards the horizon \cite{Gubser:2006bz,Herzog:2006gh,Rey:1998ik,Brandhuber:1998bs,Maldacena:1998im,Rey:1998bq}. The quark which carries a fundamental charge under the gauge group is infinitely massive and attaches on the boundary. AdS/CFT correspondence offers a different perspective on studying the strongly coupled QGP. One can investigate the various aspects with respect to QGP in the gravity spacetime. The unique advantage of holography is in dealing with real time non-equilibrium dynamics and many important insights in studying the different nature of strongly coupled plasma have been given from holography.

Unlike QCD, the $\mathcal{N} = 4$ SYM is a conformal, scale-invariant and supersymmetric theory. It is a significant challenge to apply the AdS/CFT in QCD. The approach of bottom-up (namely holographic QCD) can realize this aim, including hard wall \cite{Erlich:2005qh} and soft wall models \cite{Karch:2006pv,Batell:2008zm,dePaula:2008fp}. The scalar field or dilaton field coupling to gravitational action breaks the conformal symmetry. In the context of holographic QCD, some properties of QCD can be simulated in the gravitational backgrounds, such as the equation of state (EoS) \cite{Gubser:2008ny,Gubser:2008yx} and the thermodynamical properties of strongly coupled plasma\cite{Gursoy:2008bu,Gursoy:2008za,Gursoy:2009jd,Gursoy:2009kk,Noronha:2009ud}. In holographic dictionary, the time component of U(1) gauge field in the gravitational action is dual to the chemical potential in the boundary theory. The thermodynamical properties and  phase structure have been studied widely in the holographic QCD model with finite temperature and chemical potential\cite{DeWolfe:2010he,DeWolfe:2011ts,Cai:2012xh,Yang:2014bqa,Finazzo:2016psx,Knaute:2017opk,Sin:2007ze,Colangelo:2010pe,Ballon-Bayona:2020xls,He:2013qq,Yang:2015aia}.
Other interesting and important work in the holographic QCD can be seen in \cite{Mahapatra:2020wym,Dudal:2017max,Bohra:2019ebj,Arefeva:2020uec,Li:2017tdz,Chen:2019rez,Chen:2020ath,He:2020fdi,Mamani:2020pks,Arefeva:2020vae,Arefeva:2020byn,Bohra:2020qom,Chen:2021gop,Zhao:2021ogc,Zhou:2021nbp}. The effect of momentum anisotropy on QCD thermodynamics can be seen in \cite{Zhang:2020zrv}.

The hard partons \cite{Matsui:1986dk} which produced in a hard scattering of the collision are expected to provide some crucial inspiration about the entire evolution of the QGP. Studying the energy loss in the Heavy Ion Collisions at RHIC and LHC has been significative \cite{Qin:2015srf}. When the energetic parton with a large transverse momentum passing through the QGP, it can radiate gluons thereby loses energy as it interacts with the hot dense matter. One mechanism of energy loss is drag force \cite{Gubser:2006bz,Herzog:2006gh}. When the external heavy quark passing through the hot dense matter with a fixed velocity $\upsilon$, it feels a drag force. The energy loss can be determined by the loss of averaged momentum per unit time. Much work so far have focused on drag force. \cite{Matsuo:2006ws,Caceres:2006dj,Rougemont:2015wca,Cheng:2014fza,Mamo:2016xco,Zhang:2018mqt,Arefeva:2020bjk} has studied the drag force with an electromagnetic field/chemical potential. Drag force in non-relativistic theories and asymptotically Lifshitz spacetime has been studied in \cite{Akhavan:2008ep,Hartnoll:2009ns,Giataganas:2013hwa}. The effect of hyperscaling violation on drag force have been reported in \cite{Sadeghi:2014lha,Alishahiha:2012cm,Kiritsis:2012ta,Kioumarsipour:2018tmf}. The energy loss of rotating string has been discussed in \cite{Fadafan:2008bq,Athanasiou:2010pv,AliAkbari:2011ue,Fadafan:2012qu}. In \cite{Atashi:2016cso,Hou:2021own}, the energy loss of rotating quarks with angular velocity has been discussed. Drag force in Kerr black hole can be seen in \cite{NataAtmaja:2010hd,Arefeva:2020jvo}. Other important work can be seen in \cite{Gubser:2006nz,Nakano:2006js,Talavera:2006tj,Roy:2009sw,Panigrahi:2010cm,Chernicoff:2012iq,Chakraborty:2014kfa,Zhang:2019cxu,Andreev:2017bvr}.

In the present work, we study the thermodynamics of QCD in the D dimensions background. It is found that the scalar field $\phi(z)$ satisfies the boundary conditions and scalar field is real in the bulk. From the potential reconstruction method, the black hole solutions satisfy the Breitenlohner-Freedman (BF) bound which implies the gravitational background is stable. Also, the scalar potential is bounded from above by its UV boundary value.

The black hole thermodynamics and equations of state (EoS) have been discussed in D dimensions background. We study the Hawking temperature and free energy in the small and large black holes. From the results, a phase transition from small black hole to large black hole may exist when increasing the temperature $T$ at $0<\mu<\mu_c$ region. When $\mu\geq\mu_c$, the unstable black hole disappears. From the phase diagram, we find the phase transition temperature decreases with chemical potential which is consistent with the finding from the lattice \cite{Bellwied:2015rza}. It indicates that this holographic construction in the D dimensions provides a self-consistent framework to study the strong coupled plasma at finite temperature and chemical potential. One can observe that baryon density $\rho$ and entropy increase with the temperature in stable black hole while decreases with $T$ in unstable branch. Moreover, we also study the squared speed of sound in the holographic QCD model. The positive/negative regions of $C^2_s$ correspond to the dynamical stability/instability when $0<\mu<\mu_c$. From the results of trace anomaly, it is found that the peak of trace anomaly increases with the chemical potential which is consistent with the lattice results \cite{Borsanyi:2012cr}. One can find the equations of state are multi-valued near the phase transition temperature when $0<\mu<\mu_c$ while is single-valued when $\mu\geq\mu_c$ in different dimensions.

Furthermore, we study the temperature and chemical potential dependence of the drag force and diffusion coefficient in this holographic QCD model. The energy loss of heavy quark shows an enhancement near the phase transition temperature. This finding illustrates that the heavy quark energy loss has a nontrivial and non-monotonic dependence on temperature in D dimensions. Similar phenomenon also has been found from the results of diffusion coefficient. Furthermore, we find the heavy quark may lose less energy in higher dimension. The diffusion coefficient is larger in higher dimension.

This paper is organized as follows. In Sec.~\ref{sec:02}, we introduce the background geometry of the Einstein-Maxwell-scalar gravity system in D spacetime dimensions. In Sec.~\ref{sec:03}, we investigate the black hole thermodynamics and  equations of state (EoS) in D dimensions. In Sec.~\ref{sec:04}, we study the enhancement of energy loss around phase transition in the holographic QCD model. In Sec.~\ref{sec:05}, we give some conclusion and discussion.
\section{Background geometry}\label{sec:02}

The Einstein-Maxwell-scalar (EMs) gravity system has been extensively studied in the construction of bottom-up holographic QCD models. In this section, we review the main points of the derivation presented in \cite{Mahapatra:2020wym} at finite temperature and chemical potential in D dimensions. In order to study the Einstein-Maxwell-scalar background in D spacetime dimensions, one can consider the following Einstein-Maxwell-scalar action
\begin{equation}
\label{eqa}
 \ S = -\frac{1}{16\pi G_D }\int d^D x \sqrt{-g} \left [R-\frac{f(\phi)}{4}F_{MN}F^{MN}-\frac{1}{2}\partial_M \phi \partial^M \phi-V(\phi) \right],
\end{equation}
where $\phi$ is the scalar field and $V(\phi)$ denotes the potential of the scalar field. $f(\phi)$ is the gauge kinetic function which represents the coupling between $U(1)$ gauge field $A_M$ and scalar field. $F_{MN}$ represents the field strength tensor of the gauge field. ${G_D}$ is the corresponding D-dimensional Newton constant. We set ${G_D}$ to one in numerical calculations.

In order to investigate thermodynamical properties of QCD plasma and energy loss of quarks in the gravity background, one can take the following $Ans\ddot{a}tze$ of the background metric with a planar horizon in the Einstein frame of \cite{Mahapatra:2020wym}
\begin{equation}
\label{eqb}
\begin{split}
& ds^{2}=\frac{L^2 e^{2 P(z)}}{z^2} \left[ -g(z)dt^2+\sum^{D-2}_{i=1} dx_{i}^{2}+\frac{dz^{2}}{g(z)} \right ],\\
& \phi=\phi(z),\ A_{M}=A_t (z)\delta^t_M,
 \end{split}
\end{equation}
where $z$ is the holographic radial coordinate and it runs from $z=0$ (corresponds to the asymptotic boundary where the strongly coupled gauge theories are located) to $z=z_h$ (corresponds to the radial location of the black hole horizon). Here, $g(z)$ denotes the blackening function and $L$ represents the AdS length scale. We set $L$ to one in numerical calculations. P(z) represents the scale factor and the explicit form of $\phi$ depends on P(z).

With the $Ans\ddot{a}tze$ of Eq.(\ref{eqb}) one can get there Einstein equations of motion
\begin{equation}
\label{eqc}
 \  g''(z)+(D-2)g'(z) \left( P'(z)-\frac{1}{z} \right) -\frac{e^{-2 P(z)}z^2 f(z)A'_{t}(z)^2}{L^2} = 0,
\end{equation}

\begin{equation}
\label{eqd}
 \ P''(z)-P'(z)\left( P'(z)-\frac{2}{z} \right) +\frac{\phi'(z)^2}{2(D-2)}= 0,
\end{equation}

\begin{equation}
\label{eqe}
\begin{split}
 \ & \frac{g''(z)}{4g(z)}+\frac{D-2}{2} P''(z)+(D-2)^2 P'(z) \left(-\frac{1}{z}+\frac{P'(z)}{2}+\frac{3}{4(D-2)}\frac{g'(z)}{g(z)}\right)\\
 &-\frac{3(D-2)}{4}
 \frac{g'(z)}{z g(z)}+ \frac{e^{2 P(z)L^2 V(z)}}{2z^2 g(z)}+\frac{(D-1)(D-2)}{2z^2} = 0.
   \end{split}
\end{equation}

Similarly, one can obtain the equation of motion for the gauge field
\begin{equation}
\label{eqg}
 \ A''_t (z)-A'_t (z)\left(\frac{f'(z)}{f(z)}+(D-4)P'(z)-\frac{D-4}{z} \right) = 0,
\end{equation}
and the equation of motion for the  scalar field
\begin{equation}
\label{eqf}
\begin{split}
 \ &\phi''(z)+\phi'(z)\left(\frac{g'(z)}{g(z)}+(D-2)P'(z)-\frac{D-2}{z}\right)\\
 &+\frac{e^{-2P(z)}z^2 A'_t (z)^2}{2L^2 g(z)}\frac{\partial f(\phi)}{\partial\phi} -\frac{L^2 e^{2 P(z)}}{z^2 g(z)} \frac{\partial V(\phi)}{\partial\phi} = 0.
   \end{split}
\end{equation}

One can treat Eq.(\ref{eqf}) as a constrained equation and consider Eq.(\ref{eqc})-(\ref{eqg}) as independent. One can impose the following boundary conditions to solve the independent equations
\begin{equation}
\label{eqh}
\begin{split}
& g(0)=1,\   g(z_h)=0,\\
& A_t(0)=\mu, \    A_t(z_h)=0,\\
& P(0)=0,\ \phi(0)=0,
 \end{split}
\end{equation}
where $\mu$ is the chemical potential which is related to the near boundary expansion of the zeroth component of the gauge field from the holography. Moreover, the scalar field $\phi$ is demanded to be real in the bulk.

By solving Eq.(\ref{eqg}) and using the boundary conditions above, one can obtain the solution for the gauge field $A_t(z)$
\begin{equation}
\label{eqi}
 \ A_t (z) = \widetilde{\mu} \int^{z_h}_z d\xi \frac{e^{-(D-4)P(\xi)}\xi^{D-4}}{f(\xi)}.
\end{equation}

Using the holographic dictionary one can obtain the vacuum expectation value of the baryon density $\rho$ from the gauge field when close to the boundary, $A_t=\mu- \rho z^{D-3}$.

Then plugging Eq.(\ref{eqi}) into Eq.(\ref{eqc}), one can get the solution of $g(z)$
\begin{equation}
\label{eqj}
\begin{split}
 \ &g (z) = 1+ \int^{z}_0 d \xi e^{-(D-2)P(\xi)} \xi^{D-2} [ C_{1}+K(\xi) ],\\
 & C_{1}=-\frac{1+\int^{z_h}_0 d\xi e^{-(D-2)P(\xi)} \xi^{D-2} K(\xi)}{\int^{z_h}_0 d\xi e^{-(D-2)P(\xi)}\xi^{D-2}},\ K(\xi)=\int d\xi \left[\widetilde{\mu}^2 \frac{\xi^{D-4}e^{-(D-4)P(\xi)}}{L^2 f(\xi)} \right].
  \end{split}
\end{equation}

The expression of scalar field can be obtained in terms of $P(z)$ from Eq.(\ref{eqd})
\begin{equation}
\label{eqk}
 \ \phi(z)= \int dz \sqrt{2(D-2)\left[-P''(z)+P'(z)(P'(z)-\frac{2}{z})\right]}+C_{2},
\end{equation}
where the constant $C_{2}$ is used to ensure $\phi$ vanishes near the asymptotic boundary.

The potential of scalar field $V(\phi)$ can be solved from Eq.(\ref{eqe})
\begin{equation}
\label{eql}
 \begin{split}
V(z) & = -\frac{2z^{2}g(z)e^{-2P(z)}}{L^2} [\frac{(D-1)(D-2)}{2z^2}-\frac{3(D-2)}{4}\frac{g'(z)}{zg(z)}+\frac{g''(z)}{4g(z)} \\
 &+\frac{D-2}{2}P''(z)
+(D-2)^2 \left(-\frac{1}{z}+\frac{P'(z)}{2}+\frac{3}{4(D-2)}\frac{g'(z)}{g(z)}\right)P'(z)].
 \end{split}
\end{equation}

From Eq.(\ref{eqi})-(\ref{eql}), it is obvious that the nontrivial inputs which need to be fixed are scale factor $P(z)$ and gauge kinetic function $f(z)$. Indeed, Eq.(\ref{eqi})-(\ref{eql}) are a gravity solution for EMs system in terms of arbitrary $P(z)$ and $f(z)$. Nonetheless, one can constrain $P(z)$ and $f(z)$ to match the properties of the boundary strongly coupled gauge theory with real QCD such as linear Regge trajectory, confinement/deconfinement phase transition, etc.

In this paper, we want to study the thermodynamics and energy loss around phase transition without worrying too much about the dual boundary theory. We consider a simpler form $f(z)=e^{-(D-4)P(z)}$ proposed in Ref.\cite{Mahapatra:2020wym}. One should notice that this form does not modify the asymptotic structure of the spacetime but has a better control over the integrals which appear in Eq.(\ref{eqi})-(\ref{eql}).

Then we choose the following simple form of $P(z)$ \cite{Li:2017tdz}
\begin{equation}
\label{eqn}
 \ P(z)= -a\log(b z^2 +1).
\end{equation}

In \cite{Li:2017tdz}, the authors used same form of Eq.(\ref{eqn}) but different $f(z)$ to study the phase structure of light quarks. Using Eq.(\ref{eqn}), we fix the confinement/deconfinement temperature at zero chemical potential to be $T_{\scriptscriptstyle HP} = 270 \ MeV$ for pure gluon system. Then we study the thermodynamics of QCD in D dimensional background. It is obvious that $P(z)\rightarrow 0$ at the boundary $z = 0$ and indicates the bulk spacetime asymptotes to AdS at the boundary. Note that the forms of $P(z)$ is not changeless. One can choose other forms of $P(z)$ to construct the gravity model.

We will study the thermodynamics and energy loss around the phase transition temperature in D=4 and D=5 cases. Demanding the phase transition temperature $T_{\scriptscriptstyle HP} = 270MeV$ when $\mu=0$, we fix the parameters $a=9.889, b=0.0358$ when D=4 and $a=8.675, b=0.019$ when D=5. It should be mentioned that the results of D=4 and 5 cases are just for examples. One also can study the thermodynamics and energy loss in higher dimension. In fact, the black hole solutions always satisfy the Einstein equations of motion when D=6 or D=7 in this D-dimensional model. Note that, the temperature of phase transition is not fixed in \cite{Mahapatra:2020wym}. In this work, we fix the phase transition temperature to study the equations of state and energy loss around phase transition temperature in D dimensions.

\begin{figure}[H]
    \centering
      \setlength{\abovecaptionskip}{0.1cm}
    \includegraphics[width=12cm]{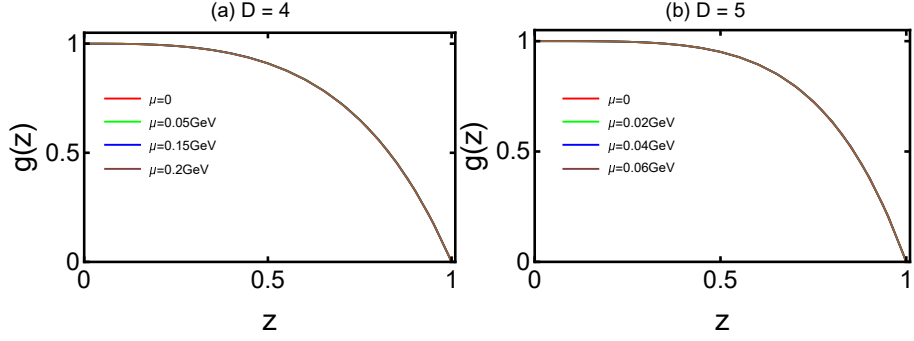}
    \caption{\label{fig1} $g(z)$ versus $z$ for different values of the chemical potential $\mu$ when $z_{h}=1$ in D=4 and 5 cases.}
\end{figure}

\begin{figure}[H]
    \centering
      \setlength{\abovecaptionskip}{0.1cm}
    \includegraphics[width=12cm]{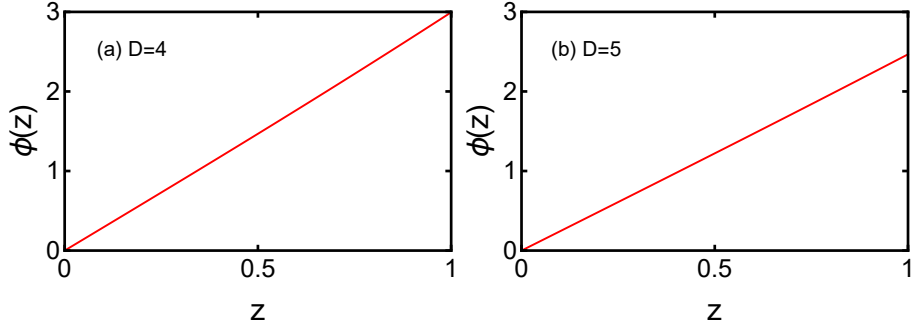}
    \caption{\label{fig2} $\phi(z)$ versus $z$ when D=4 and 5.}
\end{figure}

We plot $g(z)$ versus $z$ for different values of the chemical potential $\mu$ and $\phi(z)$ versus $z$ when D=4 and 5 in Fig.~\ref{fig1} and ~\ref{fig2}, respectively. It is found that the blackening function $g(z)$ and scalar field $\phi(z)$ satisfy the boundary conditions of Eq.(\ref{eqh}). Moreover, the scalar field is real in the bulk.

We expand the scalar field $\phi$ and scalar potential $V(\phi)$ near the asymptotic boundary $z=0$  and rewrite the scalar potential in terms of scalar field
\begin{equation}
\label{eqo}
 \begin{split}
&  D=4:\ V= -6+\frac{m^2}{2} \phi^2+\cdot\cdot\cdot , \\
 & D=5:\ V= -12+\frac{m^2}{2}\phi^2+\cdot\cdot\cdot .
 \end{split}
\end{equation}
where $m^2$ denotes the mass of the scalar field. Form Eq.(\ref{eqo}), one can get the $m^2= -2$ when D=4 and $m^2= -3$ when D=5. The results satisfy the Breitenlohner-Freedman (BF) bound \cite{Breitenlohner:1982jf}, $m^2\geq -(D-1)^2 /4$, implying that the gravitational background is stable in AdS space when D=4 and 5 .

\begin{figure}[H]
    \centering
      \setlength{\abovecaptionskip}{0.1cm}
    \includegraphics[width=14cm]{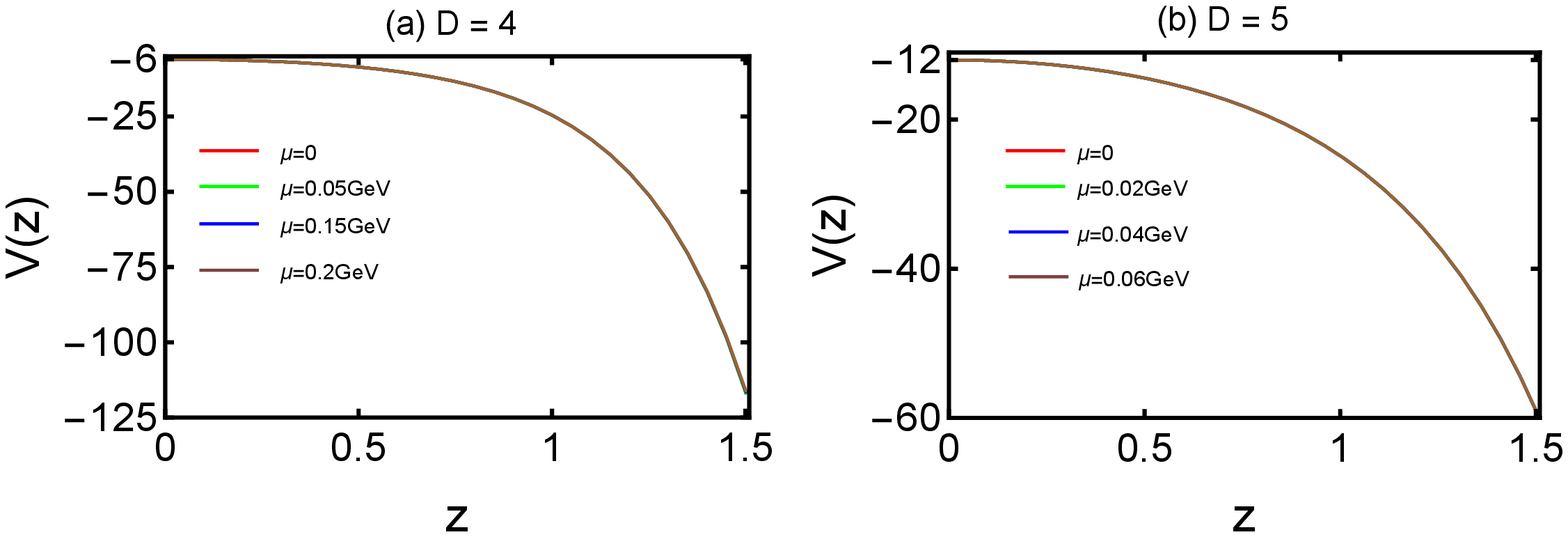}
    \caption{\label{fig3} $V(z)$ versus $z$ for different values of the chemical potential $\mu$ when $z_{h}=1.5$ in D=4 and 5 cases.}
\end{figure}

\begin{figure}[H]
    \centering
      \setlength{\abovecaptionskip}{0.1cm}
    \includegraphics[width=14cm]{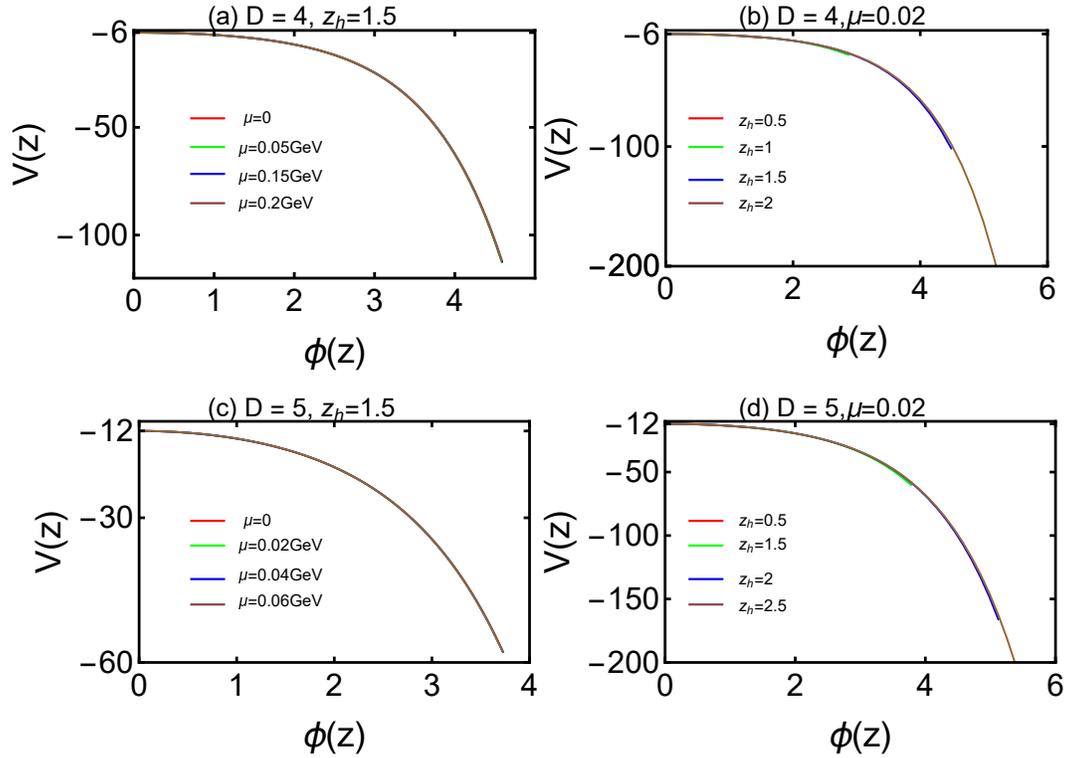}
    \caption{\label{fig4} $V(z)$ versus $\phi(z)$ when D=4 and 5.}
\end{figure}

We plot $V(z)$ versus $z$ for different values of the chemical potential when D=4 and 5 in Fig.~\ref{fig3}. We can find $V(0)\geq V(z)$ in the bulk which indicates the Gubser criterion is satisfied in this gravity system \cite{Gubser:2000nd}. It means the scalar potential is bounded from above by its UV boundary value. In Fig.~\ref{fig4}, we plot $V(z)$ versus $\phi(z)$ when D=4 and 5. It is found that $V(z)$ depends slightly on horizon $z_h$ and chemical potential $\mu$. Indeed, the scalar potential are almost indistinguishable in the region away from the horizon while cognizable slightly near the horizon.

\section{Thermodynamics in the dynamical AdS/QCD model}\label{sec:03}

The expressions of black hole entropy and Hawking temperature in D dimensions are
\begin{equation}
\label{eqm}
 \begin{split}
&  s=\frac{L^{D-2}e^{(D-2)P(z_{h})}}{4G_{D}z^{D-2}_{h}}, \\
 & T=\frac{z^{D-2}_h e^{-(D-2)P(z_{h})}}{4\pi} \left[-K(z_h)+\frac{1+\int^{z_h}_0 d\xi e^{-(D-2)P(\xi)} \xi^{D-2} K(\xi)}{\int^{z_h}_0 d\xi e^{-(D-2)P(\xi)}\xi^{D-2}}\right].
 \end{split}
\end{equation}

From the first law of thermodynamics, one can obtain the free energy at fixed chemical potential and volume

\begin{equation}
\label{eqn1}
 F=\int^\infty_{z_h} s \frac{dT}{d z_h}d z_h,
\end{equation}
where we have fixed $F(z_h \rightarrow \infty)=0$ to ensure the free energy of the thermal gas background background to be zero.

\subsection{Black hole thermodynamics}

\begin{figure}[H]
    \centering
      \setlength{\abovecaptionskip}{0.1cm}
    \includegraphics[width=14cm]{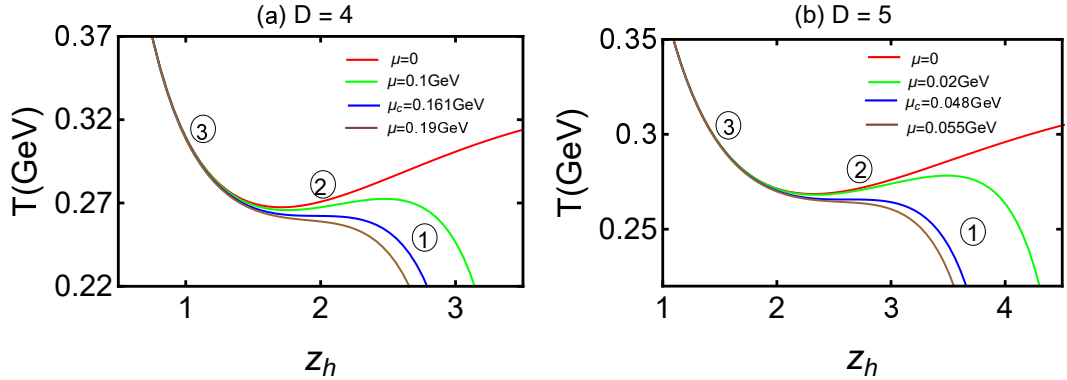}
    \caption{\label{fig5} Temperature $T$ as a function of horizon $z_{h}$ for different values of the chemical potential $\mu$ when D=4 and 5.}
\end{figure}

\begin{figure}[H]
    \centering
      \setlength{\abovecaptionskip}{0.1cm}
    \includegraphics[width=14cm]{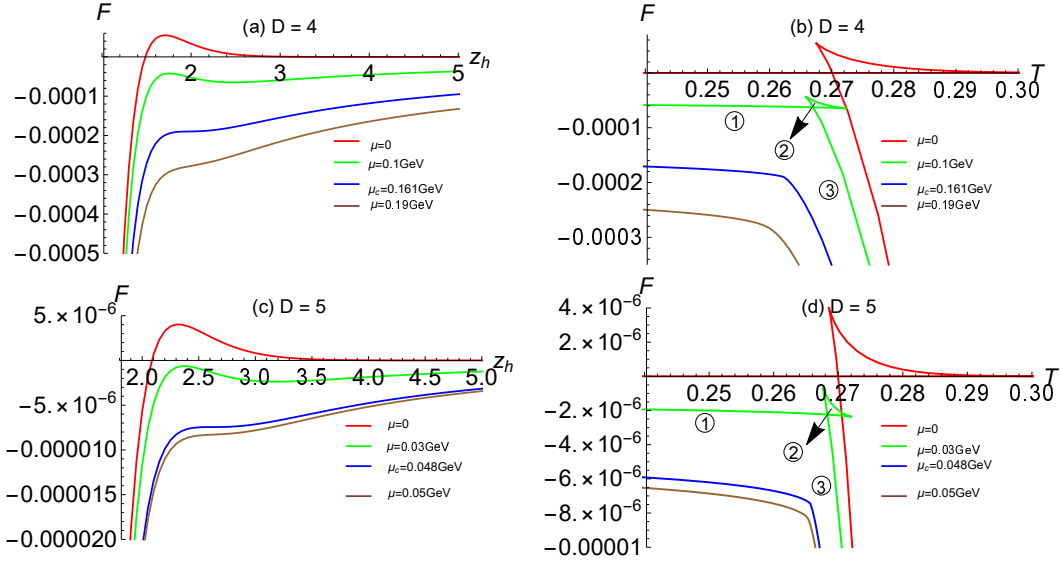}
    \caption{\label{fig6} Free energy $F$ as a function of horizon $z_{h}$ and temperature $T$ when D=4 and 5.}
\end{figure}

We plot temperature $T$ as a function of horizon $z_{h}$ for different values of the chemical potential $\mu$ when D=4 and 5 in Fig.~\ref{fig5}. It is found that there are two branches (\textcircled{2}, \textcircled{3}) in $(T, z_h)$ plane when $\mu=0$. There is a minimum temperature below which the black hole solution does not exist. Indeed, three branches (\textcircled{1}, \textcircled{2}, \textcircled{3}) exist simultaneously when $0<\mu<\mu_c$.
The small black hole \textcircled{1} (large $z_h$ region) and large black hole \textcircled{3} (small $z_h$ region) where $T$ decreases with $z_h$ are thermodynamically stable. The branch \textcircled{2} (slope is positive) for which $T$ increases with $z_h$ is unstable from the thermodynamic point of view. A phase transition from small black hole to large black hole may exist when increases the temperature $T$. When $\mu\geq \mu_c$, the unstable branch disappears and the temperature $T$ monotonously decreases with $z_h$. We also can observe this stable-unstable nature of these branches from the free energy behaviors in Fig.~\ref{fig6}.

In Fig.~\ref{fig6}, we plot free energy $F$ as a function of horizon $z_{h}$ and temperature $T$ when D=4 and 5. We first study free energy as a function of horizon in Fig.~\ref{fig6} (a) and (c). In $\mu=0$ case, one can observe that the free energy $F$ starts with a larger negative value to a positive maximum value and finally decreases to zero. There may exist a phase transition between the large black hole and thermal gas background when free energy intersects the x-axis. Indeed, the free energy has a maximum value and a minimum value simultaneously when $0<\mu<\mu_c$. This phenomenon implies a phase transition may exist between the large black hole and small black hole. When $\mu\geq \mu_c$, the free energy becomes monotonous and no large black hole to small black hole phase transition happens when increasing $z_h$. Indeed, the small-large black hole phase transition in the gravitational background is dual to the confinement-deconfinement phase transition in the boundary theory \cite{He:2013qq,Yang:2015aia}.

Then we study free energy as a function of temperature in Fig.~\ref{fig6} (b) and (d). When $\mu=0$, a phase transition between the large black hole phase and thermal-AdS phase happens at $T_{\scriptscriptstyle HP} =270MeV$. When $0<\mu<\mu_c$, a characteristic swallow-tailed structure (\textcircled{2}) appears which implies a first order phase transition happens. The phase transition occurs at the kink which represents a transition between the large black hole and small black hole. We can find that the free energy of the second branch (\textcircled{2}) is always larger than other phases which indicates the thermodynamically unstable nature of this phase. The swallow-tailed structure gradually decreases with $\mu$ and it completely disappears at $\mu= \mu_c$. When $\mu\geq \mu_c$, the free energy monotonously decreases with the temperature.
\begin{figure}[H]
    \centering
      \setlength{\abovecaptionskip}{0.1cm}
    \includegraphics[width=14cm]{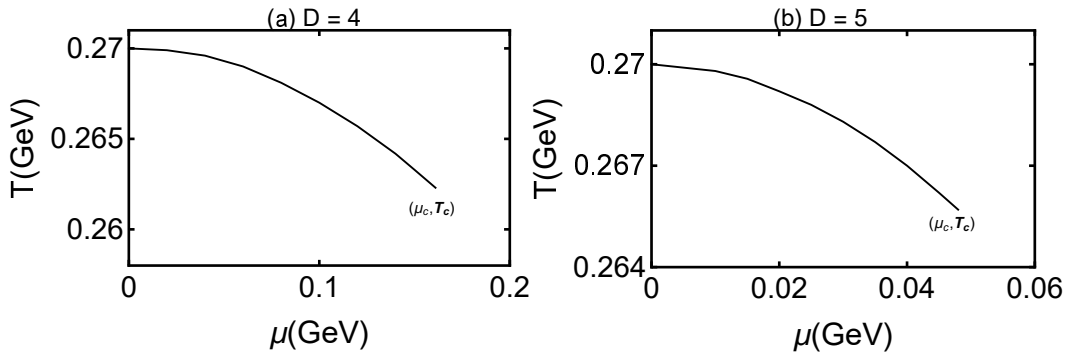}
    \caption{\label{fig7} The phase diagram in $T-\mu$ plane when D=4 and 5. The system undergoes a first-order phase transition at finite $T$ and stops at the critical point $(\mu_c,T_c)$. $(\mu_c,T_c)\simeq(0.161GeV,0.2623GeV)$ in D=4 case and $(\mu_c,T_c)\simeq(0.048GeV,0.2657GeV)$ in  D=5 case. Then the phase transition becomes a crossover when $\mu>\mu_c$ .}
\end{figure}

In Fig.~\ref{fig7}, we plot the phase diagram in $T-\mu$ plane when D=4 and 5. It is found that the
phase transition temperature decreases with $\mu$. The first order phase transition happens between large black hole and thermal gas background when $\mu=0$. When $0<\mu<\mu_c$, the first order phase transition exists between the large black hole and the small black hole. The first order phase transition terminates at $\mu= \mu_c$. When $\mu\geq \mu_c$, the system is crossover. The behavior of the phase diagram resembles the Van der Waal-like liquid-gas phase transition from the results. The phase diagram for heavy quarks and even pure gauge limit is first order, then crossover \cite{Cai:2012xh,He:2013qq,Yang:2015aia,Dudal:2017max}.

In this subsection, we study the black hole thermodynamics and phase diagram in different dimensions. In \cite{Dudal:2017max}, the authors use a different form of gauge kinetic function to match the meson mass spectrum. A different form of scale factor also has been used to fix the phase transition temperature of pure gluon system. In this work, we also fix the temperature of phase transition at zero chemical potential to be $T_{\scriptscriptstyle HP} = 270 \ MeV$ for pure gluon system. From the results, the behaviors of black hole thermodynamics and phase diagram are similar with \cite{Dudal:2017max}. In \cite{Li:2017tdz}, the authors used same form of Eq.(\ref{eqn}) but different gauge kinetic function to study the phase structure of light quarks. The behavior of phase diagram in \cite{Li:2017tdz} is different. In \cite{Li:2017tdz}, the behavior of phase diagram is crossover at low density while first order phase transition occurs at high density.

In this work, the gauge kinetic function does not modify the asymptotic structure of the spacetime to match the meson mass spectrum but has a better control over the integrals in Eq.(\ref{eqi})-(\ref{eql}). We focus the behaviors of equations of state and energy loss around phase transition temperature in different dimensions. One can find that the equations of state are multi-valued near the phase transition temperature when $0<\mu<\mu_c$ while is single-valued when $\mu\geq\mu_c$ in D dimensions. In the studying of drag force, we find the heavy quark may lose less energy in higher dimension. The diffusion coefficient is larger in higher dimension. More details are in the following sections.

\subsection{Equations of state}

\begin{figure}[H]
    \centering
      \setlength{\abovecaptionskip}{0.1cm}
    \includegraphics[width=14cm]{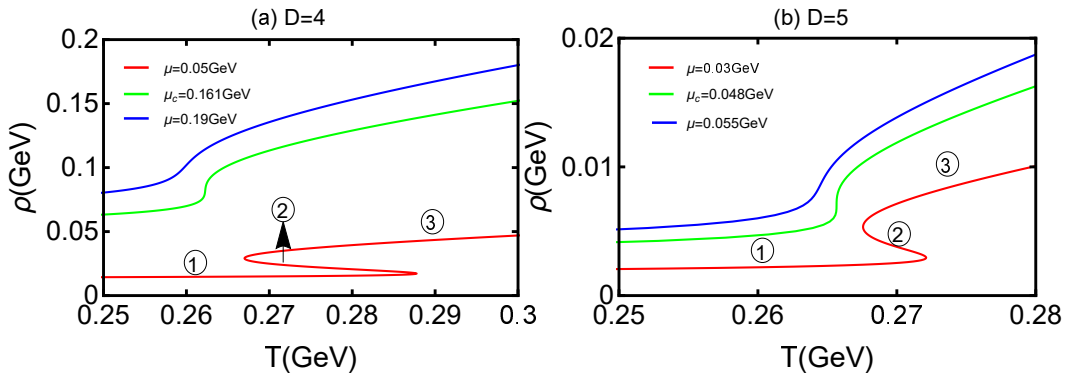}
    \caption{\label{fig8}  The baryon density $\rho$ as a function of temperature $T$ for different values of the chemical potential $\mu$ when D=4 and 5.}
\end{figure}

The baryon density $\rho$ can be obtained from the gauge field when close to the boundary, $A_t=\mu- \rho z^{D-3}$. In $D=5$ case, the expression of gauge field (Eq.(\ref{eqi})) can be obtained as $A_t= \mu- \frac{\mu}{ z^2_h} z^2$ and $\rho=\mu/ z^2_h$. In $D=4$ case, the expression of gauge field can be obtained as $A_t= \mu- \frac{\mu}{ z_h} z$ and $\rho=\mu / z_h$. In Fig.~\ref{fig8}, we plot the baryon density $\rho$ as a function of temperature $T$ for different values of the chemical potential $\mu$ when D=4 and 5. When $0<\mu<\mu_c$, $\rho$ is multi-valued and increases with the temperature in stable branches (\textcircled{1}, \textcircled{3}) while decreases with $T$ in unstable branch (\textcircled{2}). This behavior indicates a phase transition exists. When $\mu\geq \mu_c$, the baryon density becomes single-valued and invariably increases with the temperature. This implies there is no phase transition. The chemical potential increases the values of baryon density when D=4 and 5.

\begin{figure}[H]
    \centering
      \setlength{\abovecaptionskip}{0.1cm}
    \includegraphics[width=14cm]{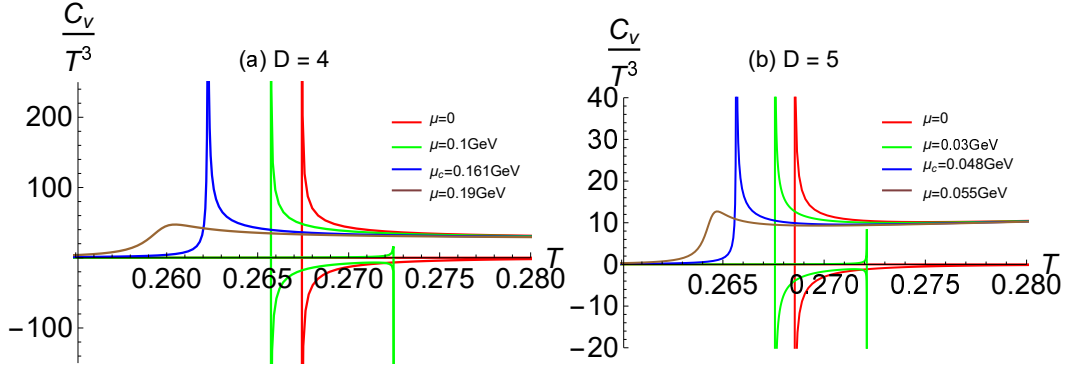}
    \caption{\label{fig9}  The specific heat $C_V/T^3$ versus temperature $T$ for different values of the chemical potential $\mu$ when D=4 and 5.}
\end{figure}

Then we study the dependence of specific heat $C_V$ on the temperature. The specific heat is defined by
\begin{equation}
\label{eqo1}
 C_V= T (\frac{\partial s}{\partial T}).
\end{equation}

In Fig.~\ref{fig9}, we plot the specific heat $C_V/T^3$ versus temperature $T$ for different values of the chemical potential $\mu$ when D=4 and 5. When $0<\mu<\mu_c$, the specific heat has negative values which corresponds to thermodynamical instability. When $\mu\geq \mu_c$, the specific heat is always positive which means the black hole is thermodynamically stable.

\begin{figure}[H]
    \centering
      \setlength{\abovecaptionskip}{0.1cm}
    \includegraphics[width=14cm]{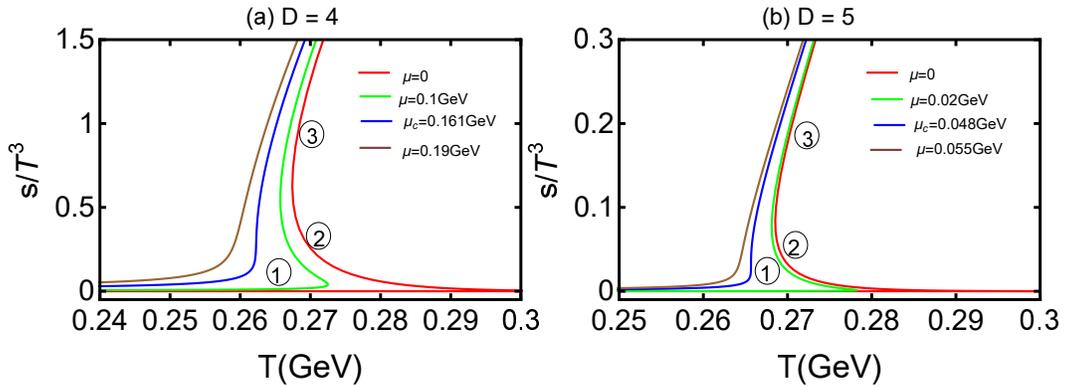}
    \caption{\label{fig10}  The entropy density $s/T^3$ versus temperature $T$ for different values of the chemical potential $\mu$ when D=4 and 5.}
\end{figure}

We plot the entropy density $s/T^3$ versus temperature $T$ for different values of the chemical potential $\mu$ when D=4 and 5 in Fig.~\ref{fig10}. When $0<\mu<\mu_c$ the entropy is multi-valued and entropy increases with the temperature in stable branches (\textcircled{1}, \textcircled{3}) while decreases with $T$ in unstable branch (\textcircled{2}). This phenomenon indicates a phase transition happens. When $\mu\geq \mu_c$, the entropy is single-valued and always increases with the temperature. No phase transition exists in this process. The chemical potential enhances the values of entropy density when D=4 and 5.

\begin{figure}[H]
    \centering
      \setlength{\abovecaptionskip}{0.1cm}
    \includegraphics[width=14cm]{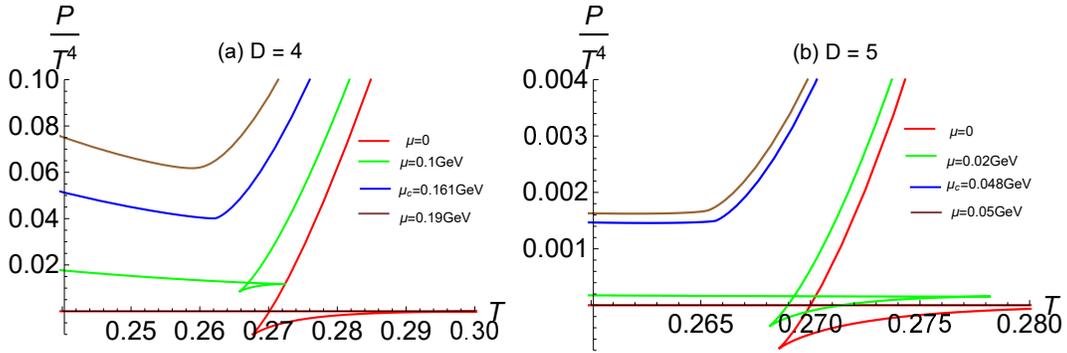}
    \caption{\label{fig11}  The pressure $p/T^4$ versus temperature $T$ for different values of the chemical potential $\mu$ when D=4 and 5.}
\end{figure}

The pressure is related to the free energy as $p=-F$. In Fig.~\ref{fig11}, we plot the pressure $p/T^4$ versus temperature $T$ for different values of the chemical potential $\mu$ when D=4 and 5. One can find the pressure increases with the chemical potential which indicates the phase transition temperature is pushed to be smaller for larger chemical potential in some sense. This phenomenon is consistent to the lattice results \cite{Borsanyi:2012cr}. The pressure is multi-valued when $0<\mu<\mu_c$ while is not when $\mu\geq\mu_c$. Moreover, the behavior of pressure for different values of the chemical potential is consistent with the results of free energy in Fig.~\ref{fig6}. Indeed, the chemical potential increases the pressure when D=4 and 5.

\begin{figure}[H]
    \centering
      \setlength{\abovecaptionskip}{0.1cm}
    \includegraphics[width=14cm]{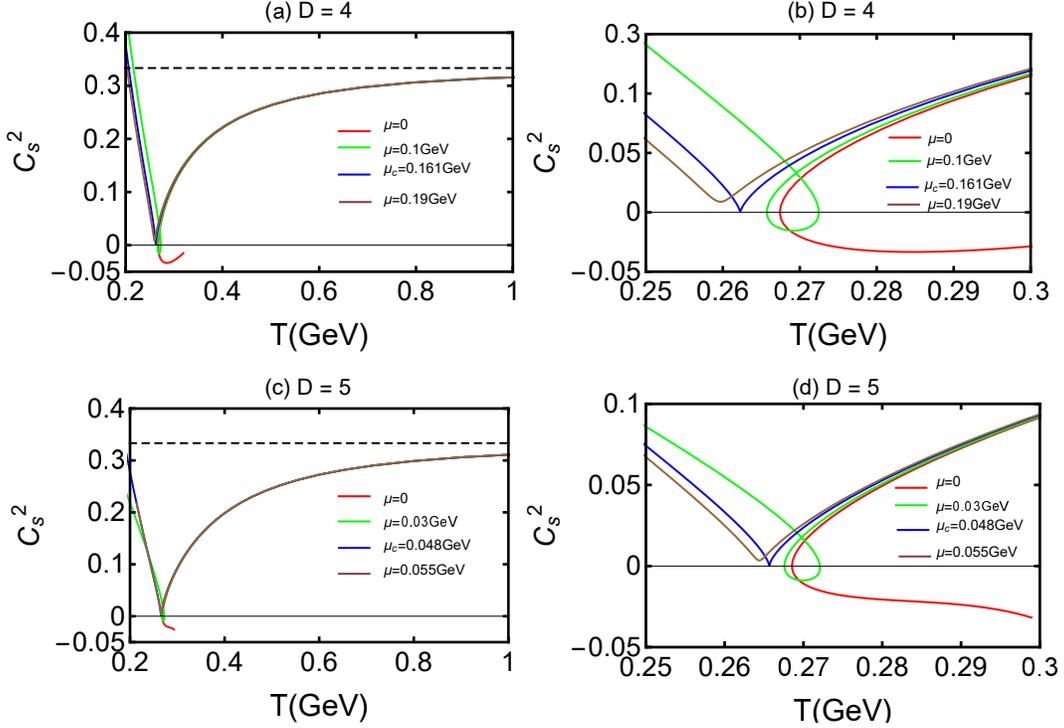}
    \caption{\label{fig12}  The squared of speed of sound $C^2_s$ versus temperature $T$ for different values of the chemical potential $\mu$ when D=4 and 5.}
\end{figure}

The squared speed of sound can be defined as
\begin{equation}
\label{eqp}
 C^2_s= \frac{\partial \ln T}{\partial \ln s}.
\end{equation}

We plot the squared of speed of sound $C^2_s$ versus temperature $T$ for different values of the chemical potential $\mu$ when D=4 and 5 in Fig.~\ref{fig12}. It is obvious that $C^2_s$
approaches to the conformal limit $1/3$ (the dotted black line) at high $T$ form Fig.~\ref{fig12}. (a) and (c). From Fig.~\ref{fig12}.(b) and (d), the positive/negative regions of $C^2_s$ correspond to the thermodynamical stability/instability when $0<\mu<\mu_c$. This result is consistent with the results of the specific heat since the negative value of specific heat implies thermodynamically unstable. In fact, the squared of speed of sound $C^2_s$ is related to the specific heat as $C^2_s= s/ C_V$. When $\mu=\mu_c$, $C^2_s = 0$ at the phase transition temperature. When $\mu>\mu_c$, $C^2_s$ is always positive and sharply decreases near the phase transition temperature. When the temperature reaches to phase transition temperature, $C^2_s$ has a minimum value. Moreover, the chemical potential reduces the values of $C^2_s$ near phase transition temperature. The influence of chemical potential on the $C^2_s$ is negligible at high temperature.

\begin{figure}[H]
    \centering
      \setlength{\abovecaptionskip}{0.1cm}
    \includegraphics[width=14cm]{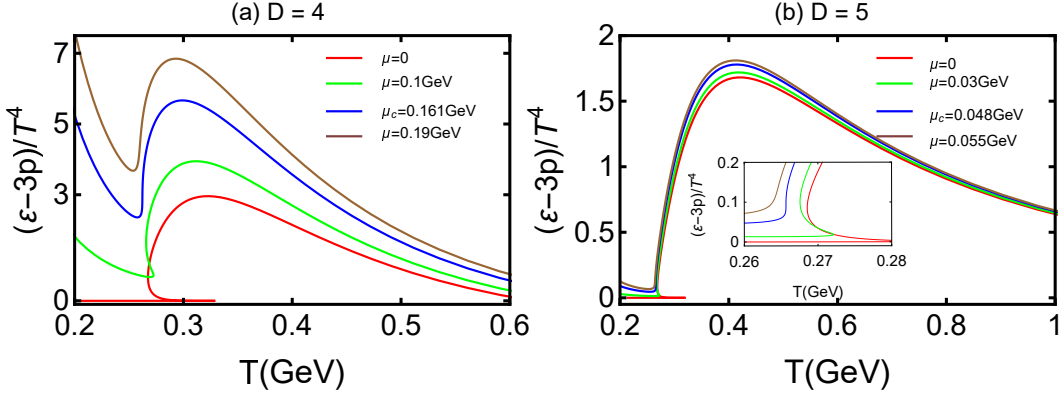}
    \caption{\label{fig13}  The trace anomaly $(\epsilon-3p)/T^4$ versus temperature $T$ for different values of the chemical potential $\mu$ when D=4 and 5.}
\end{figure}

In Fig.~\ref{fig13}, we plot the trace anomaly $(\epsilon-3p)/T^4$ versus temperature $T$ for different values of the chemical potential $\mu$ when D=4 and 5. The trace anomaly is multi-valued near the phase transition temperature when $0<\mu<\mu_c$ while is not when $\mu\geq\mu_c$. The trace anomaly has a non-monotonic dependence on temperature. In D=4 and $\mu\geq\mu_c$ case, the trace anomaly first decreases with temperature and then increases with $T$. After comes to a peak value, trace anomaly starts to decrease with temperature. However, the trace anomaly always increases with $T$ before it comes to a peak value in D=5 and $\mu\geq\mu_c$ case. It is also found that the peak value of trace anomaly increases with the chemical potential. This similar phenomenon has been found in the lattice results in \cite{Borsanyi:2012cr}. Indeed, the position of the peak shifts toward the lower temperature slightly when chemical potential increases.

In this subsection, we study the equations of state around phase transition temperature in different dimensions. One can find that the equations of state are multi-valued near the phase transition temperature when $0<\mu<\mu_c$ while is single-valued when $\mu\geq\mu_c$ in D dimensions.

\section{Nontrivial energy loss in the holographic QCD model}\label{sec:04}
In this section, we will calculate the drag force and diffusion coefficient around phase transition temperature in D spacetime dimensions. When the external heavy quark passing through the hot dense matter with a fixed velocity $\upsilon$, it feels a drag force. The energy loss can be determined by the loss of averaged momentum per unit time. It is worthy to study the energy loss of heavy quark around the phase transition temperature when heavy quark moving through the strongly coupled plasma.

\subsection{Drag force}

In trailing string model \cite{Gubser:2006bz,Herzog:2006gh}, the heavy quark passing through the hot dense medium with an invariable velocity is dual to an open string with an endpoint moving with a same speed on the boundary, while the rest of the string trailing down into the bulk of the AdS spacetime. The energy loss of the heavy quark can be seen as the energy flow (or momentum flux) from the endpoint along the string towards the horizon of the worldsheet, namely drag force. The dynamics of the heavy quarks can be described by the Brownian motion, and the equation of the motion is

\begin{equation}
\label{eqq}
 \frac{dp}{dt}=-\eta_{\scriptscriptstyle D} p+f_{1},
\end{equation}
where the drag force $f$ is equal to $-\eta_{\scriptscriptstyle D} p$. The $\eta_{\scriptscriptstyle D}$ is the drag coefficient which suppressed by the mass and the $p$ is momentum of the quark. In the condition of $\frac{dp}{dt} = 0$, the driving force $f_{1}$ is equal to the drag force $f$.

The metric of the background (\ref{eqb}) in the string frame is
\begin{equation}
\label{eqr}
ds^{2}=\frac{L^2 e^{2 P_s(z)}}{z^2} \left[ -g(z)dt^2+\sum^{D-2}_{i=1} dx_{i}^{2}+\frac{dz^{2}}{g(z)} \right ],
\end{equation}
where $P_s(z)= P(z)+ \sqrt{\frac{1}{6}} \phi(z)$.

The coordinates in Eq.(\ref{eqr}) can be parameterized by
\begin{equation}
\label{eqs}
\ t=\tau,\ \ x_{1}=vt+\xi(z),\  \ z=\sigma,
\end{equation}
where $\upsilon$ is the quark velocity.

The Lagrangian density can be obtained from the Nambu-Goto action as
\begin{equation}
\label{eqt}
\mathcal{L} =\sqrt{-g_{tt}g_{zz}-g_{zz}g_{xx}\upsilon^2-g_{tt}g_{xx}\xi'^{2}},
\end{equation}
where $g_{tt}= -\frac{L^2 e^{2 P_s(z)}}{z^2} g(z)$, $g_{xx}=\frac{L^2 e^{2 P_s(z)}}{z^2}$ and $g_{zz}=\frac{L^2 e^{2 P_s(z)}}{z^2} \frac{1}{g(z)}$.

The Lagrangian density does not depend on $\xi$ from Eq.(\ref{eqt}), which means the canonical momentum is conserved

\begin{equation}
\label{equ}
\ \Pi_{\xi}=\frac{\partial \mathcal{L}}{\partial\xi'}=\frac{-g_{tt}g_{xx} \xi'}{\sqrt{-g_{tt}g_{zz}-g_{zz}g_{xx}\upsilon^2-g_{tt}g_{xx}\xi'^{2}}}.
\end{equation}

Then one can get
\begin{equation}
\label{eqv}
\xi'^{2}=\frac{-g_{zz}(g_{tt}+g_{xx} \upsilon^2)\Pi^2_{\xi}}{g_{tt}g_{xx}(g_{tt}g_{xx}+\Pi^2_{\xi})}.
\end{equation}

Both the numerator and the denominator must change sign at a same location $z$ from Eq. (\ref{eqv}). The critical point $z_{c}$ can be written as
\begin{equation}
\label{eqw}
\ g_{tt}(z_{c})=-g_{xx}(z_{c})\upsilon^2,
\end{equation}
and
\begin{equation}
\label{eqx}
\ \Pi^2_{\xi}=-g_{tt}(z_{c})g_{xx}(z_{c}).
\end{equation}

Then the drag force can be obtained from Eq. (\ref{eqw}) and Eq. (\ref{eqx})
\begin{equation}
\label{eqy}
f=-\frac{1}{2\pi\alpha'} \Pi_{\xi}=-\frac{1}{2\pi\alpha'} g_{xx}(z_{c}) \upsilon,
\end{equation}
where the minus sign means the direction of motion is against the drag force from the medium.

In \cite{Gubser:2006bz}, the drag force of the $\mathcal{N} = 4$ SYM theory with zero chemical potential is
\begin{equation}
\label{eqz}
f_{\scriptscriptstyle SYM}= -\frac{\pi T^{2} \sqrt{\lambda}}{2} \frac{\upsilon}{\sqrt{1-\upsilon^{2}}},
\end{equation}
where $ \sqrt{\lambda}=\frac{L^{2}}{\alpha'}=\sqrt{g^{2}_{YM}N_{c}}$.

\begin{figure}[H]
    \centering
      \setlength{\abovecaptionskip}{0.1cm}
    \includegraphics[width=14cm]{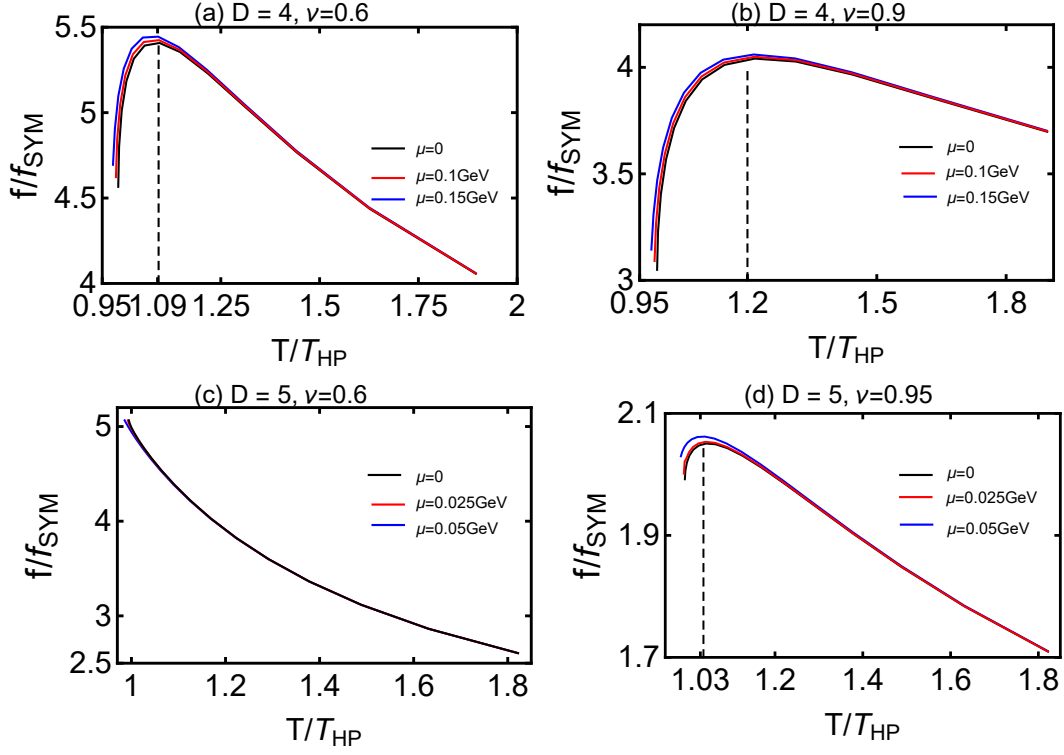}
    \caption{\label{fig14} Drag force normalized by its conformal limit $f/f_{\scriptscriptstyle SYM}$ versus temperature $T/T_{\scriptscriptstyle HP}$ for different velocity and chemical potential $\mu$ in large black hole when D=4, 5. We take $T_{\scriptscriptstyle HP}=0.27 GeV$ when $\mu=0$ in this figure.}
\end{figure}

In Fig.~\ref{fig14}, we plot drag force normalized by its conformal limit $f/f_{\scriptscriptstyle SYM}$ versus temperature $T/T_{\scriptscriptstyle HP}$ for different velocity and chemical potential $\mu$ in large black hole (high temperature and thermodynamically stable region) when D=4, 5. It is obvious that the drag force is sensitive to the quark velocity and the temperature. The energy loss of heavy quark shows an enhancement near the phase transition temperature. In D= 4 case, the peak is around $T=1.09 \ T_{\scriptscriptstyle HP}$ when $\nu = 0.6$ and the peak is around $T=1.2 \ T_{\scriptscriptstyle HP}$ when $\nu = 0.9$. This finding illustrates that the heavy quark energy loss has a nontrivial and non-monotonic dependence on temperature. It should be mentioned that the energy loss also has an enhancement even at low velocity when D= 4 which is different from \cite{Rougemont:2015wca}. In D= 5 case, a peak only appears at high velocity ($\nu = 0.95$) near $T=1.037 \ T_{\scriptscriptstyle HP}$. Furthermore, the peak value increases with the chemical potential which implies $\mu$ enhances the energy loss. The peak value of energy loss is moving towards lower temperature slightly when increasing chemical potential from the numerical results.

\begin{figure}[H]
    \centering
      \setlength{\abovecaptionskip}{0.1cm}
    \includegraphics[width=14cm]{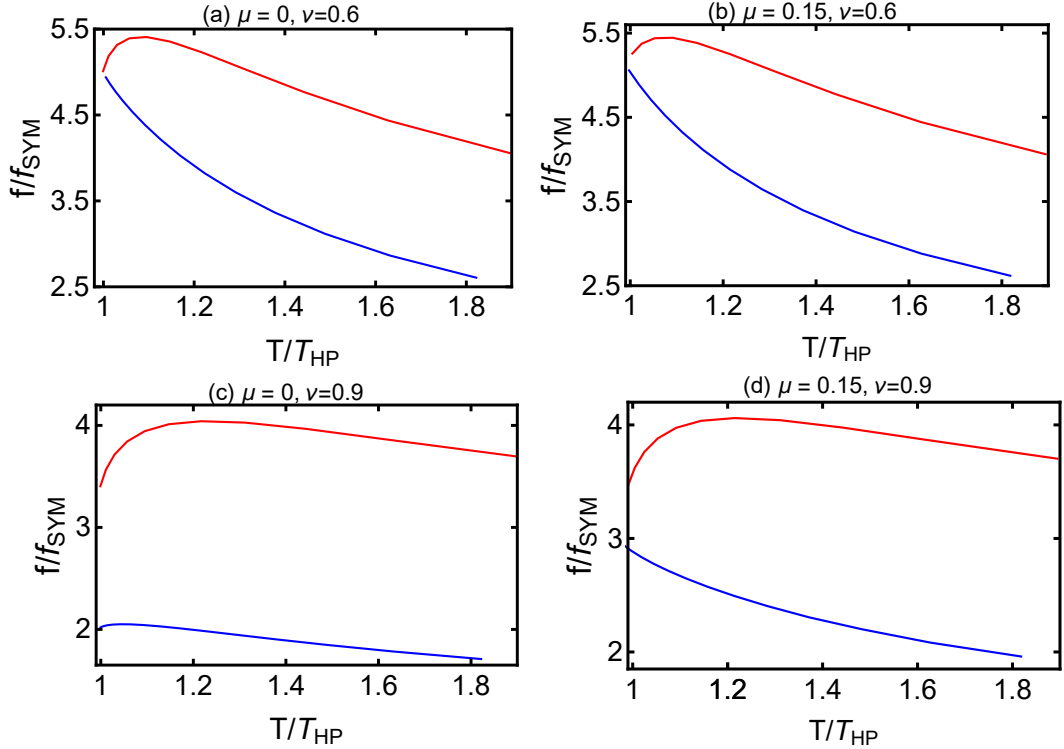}
    \caption{\label{fig15} Drag force normalized by its conformal limit $f/f_{\scriptscriptstyle SYM}$ versus temperature $T/T_{\scriptscriptstyle HP}$ for different dimensions in large black hole. We take $T_{\scriptscriptstyle HP}=0.27 GeV$ when $\mu=0$ in this figure. The red line for D=4 case and blue line for D=5 case.}
\end{figure}

In Fig.~\ref{fig15}, we plot drag force normalized by its conformal limit $f/f_{\scriptscriptstyle SYM}$ versus temperature $T/T_{\scriptscriptstyle HP}$ for different dimensions in large black hole. One can find the heavy quark may lose less energy in higher dimension.

\subsection{Diffusion coefficient}

The diffusion coefficient can be used to describe how strongly the heavy quark couples to the hot dense medium. The diffusion coefficient in the $\mathcal{N} = 4$ SYM situation (with zero chemical potential) can be given by
\begin{equation}
\label{eqz1}
D_{\scriptscriptstyle SYM}=\frac{T}{m} t_{\scriptscriptstyle SYM}=\frac{2}{\pi T \sqrt{\lambda}}.
\end{equation}
where $t_{\scriptscriptstyle SYM}= \frac{1}{\eta_{D_{\scriptscriptstyle SYM}}}$ is the diffusion time.

The Eq.(\ref{eqy}) can be rewritten as
\begin{equation}
\label{eqz2}
f= -\eta_{\scriptscriptstyle D}p,
\end{equation}
where $\eta_{\scriptscriptstyle D}$ and $p$ is the drag coefficient and the momentum, respectively.

From Eq.(\ref{eqy}), Eq.(\ref{eqz1}) and Eq.(\ref{eqz2}), diffusion coefficient normalized by the $\mathcal{N} = 4$ SYM result is
\begin{equation}
\label{eqz3}
\frac{D}{D_{\scriptscriptstyle SYM}}= \frac{\pi^2 T^2}{g_{xx}(z_{c})\sqrt{1-\upsilon^2}}.
\end{equation}

\begin{figure}[H]
    \centering
      \setlength{\abovecaptionskip}{0.1cm}
    \includegraphics[width=14cm]{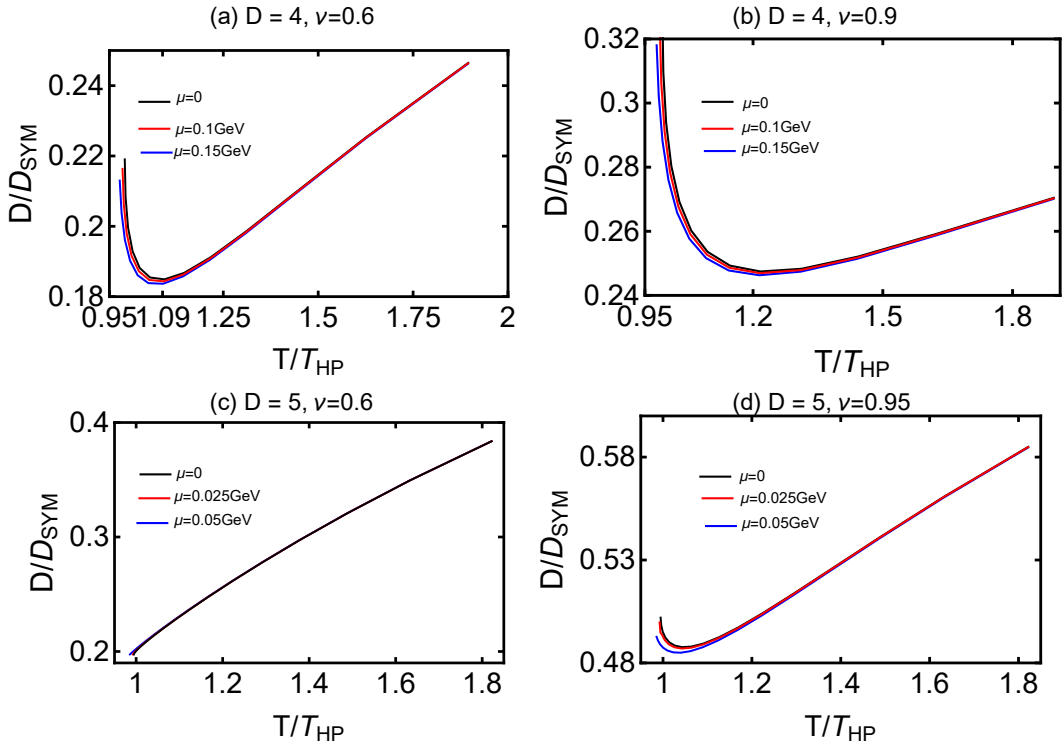}
    \caption{\label{fig16} Diffusion coefficient normalized by its conformal limit $D/D_{\scriptscriptstyle SYM}$ versus temperature $T/T_{\scriptscriptstyle HP}$ for different velocity and chemical potential $\mu$ in large black hole (high temperature region) when D=4, 5. We take $T_{\scriptscriptstyle HP}=0.27 GeV$ when $\mu=0$ in this figure.}
\end{figure}

In Fig.~\ref{fig16}, we plot diffusion coefficient normalized by its conformal limit $D/D_{\scriptscriptstyle SYM}$ versus temperature $T/T_{\scriptscriptstyle HP}$ for different velocity and chemical potential $\mu$ in large black hole when D=4, 5. Similarly to the results of drag force, the diffusion coefficient also has a nontrivial and non-monotonic temperature dependence on temperature in D dimensions. A suppression appears around the phase transition. Indeed, the diffusion coefficient decreases with the chemical potential.

\begin{figure}[H]
    \centering
      \setlength{\abovecaptionskip}{0.1cm}
    \includegraphics[width=14cm]{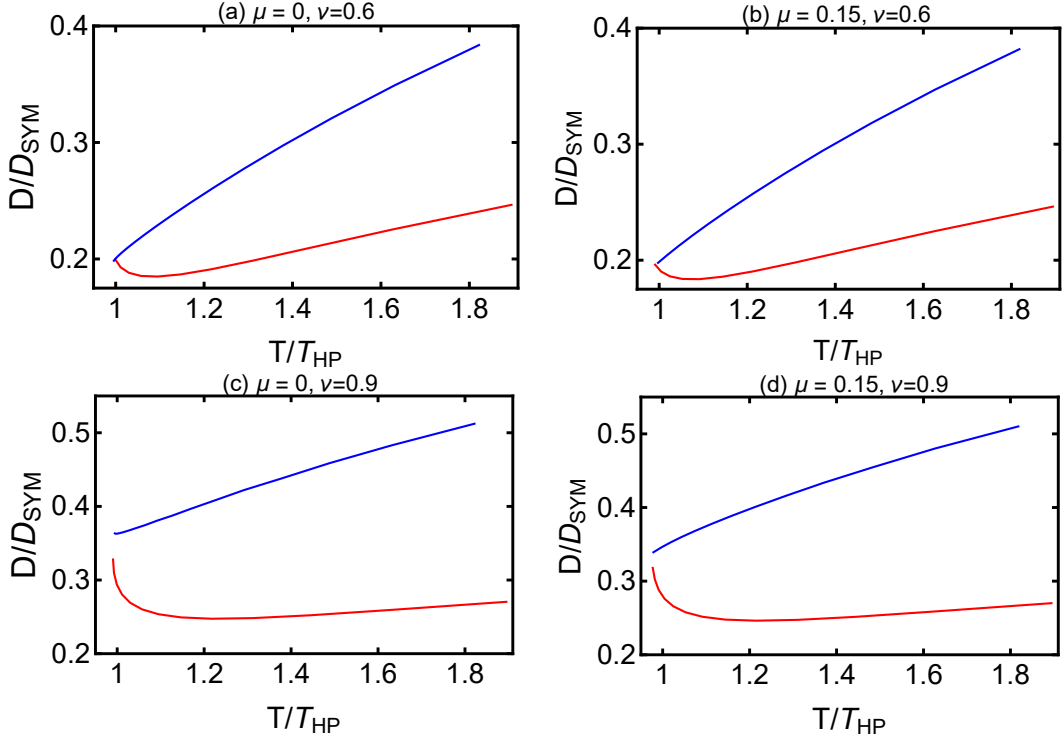}
    \caption{\label{fig17} Diffusion coefficient normalized by its conformal limit $D/D_{\scriptscriptstyle SYM}$ versus temperature $T/T_{\scriptscriptstyle HP}$ for different dimensions in large black hole (high temperature region). We take $T_{\scriptscriptstyle HP}=0.27 GeV$ when $\mu=0$ in this figure. The red line for D=4 case and blue line for D=5 case.}
\end{figure}

In Fig.~\ref{fig17}, we plot diffusion coefficient normalized by its conformal limit $D/D_{\scriptscriptstyle SYM}$ versus temperature $T/T_{\scriptscriptstyle HP}$ for different dimensions in large black hole. We find that the diffusion coefficient is larger in higher dimension.

\section{Conclusion and discussion}\label{sec:05}

In this paper, we consider the holographic QCD model with a planar horizon in the D dimensions with   different consistent metric solutions. Furthermore, we study the black hole thermodynamics, phase diagram and equations of state (EoS) and the energy loss around the phase transition temperature.

We study the Hawking temperature and free energy in the small and large black hole. One can observe the stable-unstable nature of black hole branches from the Hawking temperature and free energy behaviors. From the results, a phase transition from small black hole to large black hole may exist as one increases the temperature $T$ when $0<\mu<\mu_c$. When $\mu\geq\mu_c$, the unstable black hole disappears. From the results of phase diagram, we find the phase transition temperature decreases with chemical potential which consistent with the finding from the lattice results \cite{Bellwied:2015rza}. It is found that baryon density $\rho$ and entropy increase with the temperature in stable black hole while decreases with $T$ in unstable branch. From the studying of the speed of sound, one can observe that the positive/negative regions of $C^2_s$ correspond to the dynamical stability/instability when $0<\mu<\mu_c$. One also can find that the peak of trace anomaly increases with the chemical potential which is consistent with the lattice results in \cite{Borsanyi:2012cr}.

Furthermore, we study the temperature and chemical potential dependence of the drag force and diffusion coefficient in this holographic QCD model. It is obvious that the drag force is sensitive to the quark velocity and the temperature. The energy loss of heavy quark shows an enhancement near the phase transition temperature. This finding illustrates that the heavy quark energy loss has a nontrivial and non-monotonic dependence on temperature in D dimensions. The peak of energy loss is moving towards lower temperature slightly when increasing chemical potential. The diffusion coefficient also has a nontrivial and non-monotonic temperature dependence on temperature in D dimensions. Moreover, we find the heavy quark may lose less energy in higher dimension. The diffusion coefficient is larger in higher dimension.

We expect that the results of energy loss around phase transition temperature in this paper could provide some theoretical reference on the study of heavy quark energy loss from heavy ion collision experiments. Since a strong magnetic field has been produced in the heavy ion collisions, it is also significant to study the jet quenching parameter around phase transition temperature with nonzero chemical potential and magnetic field in the holographic QCD model. We hope report this research in the future.

\section*{Acknowledgments}

Defu Hou is supported by the NSFC Grant Nos.11735007, 11890711 and 11890710. Xian-Ming Liu is supported by the Program for Innovative Youth Research Team in University of Hubei Province of China (Grant No. T201712). We thank K. Bitaghsir Fadafan, Hai-cang Ren and Zi-qiang Zhang for valuable discussions.


\begin{thebibliography}{95}

\bibitem{Arsene:2004fa}
  I.~Arsene {\it et al.} [BRAHMS Collaboration],
  Nucl.\ Phys.\ A {\bf 757}, 1 (2005)
  doi:10.1016/j.nuclphysa.2005.02.130
  [nucl-ex/0410020].

\bibitem{Adcox:2004mh}
  K.~Adcox {\it et al.} [PHENIX Collaboration],
  Nucl.\ Phys.\ A {\bf 757}, 184 (2005)
  doi:10.1016/j.nuclphysa.2005.03.086
  [nucl-ex/0410003].

\bibitem{Back:2004je}
  B.~B.~Back {\it et al.},
  Nucl.\ Phys.\ A {\bf 757}, 28 (2005)
  doi:10.1016/j.nuclphysa.2005.03.084
  [nucl-ex/0410022].

\bibitem{Adams:2005dq}
  J.~Adams {\it et al.} [STAR Collaboration],
  Nucl.\ Phys.\ A {\bf 757}, 102 (2005)
  doi:10.1016/j.nuclphysa.2005.03.085
  [nucl-ex/0501009].

\bibitem{Aad:2013xma}
  G.~Aad {\it et al.} [ATLAS Collaboration],
  JHEP {\bf 1311}, 183 (2013)
  doi:10.1007/JHEP11(2013)183
  [arXiv:1305.2942 [hep-ex]].

\bibitem{Stephanov:2004wx}
M.~A.~Stephanov,
Prog. Theor. Phys. Suppl. \textbf{153} (2004), 139-156
doi:10.1142/S0217751X05027965
[arXiv:hep-ph/0402115 [hep-ph]].

\bibitem{Ding:2015ona}
H.~T.~Ding, F.~Karsch and S.~Mukherjee,
Int. J. Mod. Phys. E \textbf{24} (2015) no.10, 1530007
doi:10.1142/S0218301315300076
[arXiv:1504.05274 [hep-lat]].


\bibitem{Gyulassy:2004zy}
  M.~Gyulassy and L.~McLerran,
  Nucl.\ Phys.\ A {\bf 750}, 30 (2005)
  doi:10.1016/j.nuclphysa.2004.10.034
  [nucl-th/0405013].

\bibitem{Luzum:2008cw}
  M.~Luzum and P.~Romatschke,
  Phys.\ Rev.\ C {\bf 78}, 034915 (2008)
  Erratum: [Phys.\ Rev.\ C {\bf 79}, 039903 (2009)]
  doi:10.1103/PhysRevC.78.034915, 10.1103/PhysRevC.79.039903
  [arXiv:0804.4015 [nucl-th]].

\bibitem{Ryu:2015vwa}
  S.~Ryu, J.-F.~Paquet, C.~Shen, G.~S.~Denicol, B.~Schenke, S.~Jeon and C.~Gale,
  Phys.\ Rev.\ Lett.\  {\bf 115}, no. 13, 132301 (2015)
  doi:10.1103/PhysRevLett.115.132301
  [arXiv:1502.01675 [nucl-th]].

\bibitem{Gale:2012rq}
  C.~Gale, S.~Jeon, B.~Schenke, P.~Tribedy and R.~Venugopalan,
  Phys.\ Rev.\ Lett.\  {\bf 110}, no. 1, 012302 (2013)
  doi:10.1103/PhysRevLett.110.012302
  [arXiv:1209.6330 [nucl-th]].

\bibitem{Policastro:2001yc}
  G.~Policastro, D.~T.~Son and A.~O.~Starinets,
  Phys.\ Rev.\ Lett.\  {\bf 87}, 081601 (2001)
  doi:10.1103/PhysRevLett.87.081601
  [hep-th/0104066].

\bibitem{Buchel:2003tz}
  A.~Buchel and J.~T.~Liu,
  Phys.\ Rev.\ Lett.\  {\bf 93}, 090602 (2004)
  doi:10.1103/PhysRevLett.93.090602
  [hep-th/0311175].

\bibitem{Kovtun:2004de}
  P.~Kovtun, D.~T.~Son and A.~O.~Starinets,
  Phys.\ Rev.\ Lett.\  {\bf 94}, 111601 (2005)
  doi:10.1103/PhysRevLett.94.111601
  [hep-th/0405231].

\bibitem{Witten:1998qj}
  E.~Witten,
  Adv.\ Theor.\ Math.\ Phys.\  {\bf 2}, 253 (1998)
  doi:10.4310/ATMP.1998.v2.n2.a2
  [hep-th/9802150].

\bibitem{Gubser:1998bc}
  S.~S.~Gubser, I.~R.~Klebanov and A.~M.~Polyakov,
  Phys.\ Lett.\ B {\bf 428}, 105 (1998)
  doi:10.1016/S0370-2693(98)00377-3
  [hep-th/9802109].

\bibitem{Maldacena:1997re}
  J.~M.~Maldacena,
  Int.\ J.\ Theor.\ Phys.\  {\bf 38}, 1113 (1999)
  [Adv.\ Theor.\ Math.\ Phys.\  {\bf 2}, 231 (1998)]
  doi:10.1023/A:1026654312961, 10.4310/ATMP.1998.v2.n2.a1
  [hep-th/9711200].

\bibitem{Gubser:2006bz}
  S.~S.~Gubser,
  Phys.\ Rev.\ D {\bf 74}, 126005 (2006)
  doi:10.1103/PhysRevD.74.126005
  [hep-th/0605182].

\bibitem{Herzog:2006gh}
  C.~P.~Herzog, A.~Karch, P.~Kovtun, C.~Kozcaz and L.~G.~Yaffe,
  JHEP {\bf 0607}, 013 (2006)
  doi:10.1088/1126-6708/2006/07/013
  [hep-th/0605158].

\bibitem{Rey:1998ik}
  S.~J.~Rey and J.~T.~Yee,
  Eur.\ Phys.\ J.\ C {\bf 22}, 379 (2001)
  doi:10.1007/s100520100799
  [hep-th/9803001].

\bibitem{Brandhuber:1998bs}
  A.~Brandhuber, N.~Itzhaki, J.~Sonnenschein and S.~Yankielowicz,
  Phys.\ Lett.\ B {\bf 434}, 36 (1998)
  doi:10.1016/S0370-2693(98)00730-8
  [hep-th/9803137].

\bibitem{Maldacena:1998im}
  J.~M.~Maldacena,
  Phys.\ Rev.\ Lett.\  {\bf 80}, 4859 (1998)
  doi:10.1103/PhysRevLett.80.4859
  [hep-th/9803002].

\bibitem{Rey:1998bq}
  S.~J.~Rey, S.~Theisen and J.~T.~Yee,
  Nucl.\ Phys.\ B {\bf 527}, 171 (1998)
  doi:10.1016/S0550-3213(98)00471-4
  [hep-th/9803135].


\bibitem{Erlich:2005qh}
J.~Erlich, E.~Katz, D.~T.~Son and M.~A.~Stephanov,
Phys. Rev. Lett. \textbf{95} (2005), 261602
doi:10.1103/PhysRevLett.95.261602
[arXiv:hep-ph/0501128 [hep-ph]].

\bibitem{Karch:2006pv}
A.~Karch, E.~Katz, D.~T.~Son and M.~A.~Stephanov,
Phys. Rev. D \textbf{74} (2006), 015005
doi:10.1103/PhysRevD.74.015005
[arXiv:hep-ph/0602229 [hep-ph]].

\bibitem{Batell:2008zm}
B.~Batell and T.~Gherghetta,
Phys. Rev. D \textbf{78} (2008), 026002
doi:10.1103/PhysRevD.78.026002
[arXiv:0801.4383 [hep-ph]].

\bibitem{dePaula:2008fp}
W.~de Paula, T.~Frederico, H.~Forkel and M.~Beyer,
Phys. Rev. D \textbf{79} (2009), 075019
doi:10.1103/PhysRevD.79.075019
[arXiv:0806.3830 [hep-ph]].

\bibitem{Gubser:2008ny}
S.~S.~Gubser and A.~Nellore,
Phys. Rev. D \textbf{78} (2008), 086007
doi:10.1103/PhysRevD.78.086007
[arXiv:0804.0434 [hep-th]].

\bibitem{Gubser:2008yx}
S.~S.~Gubser, A.~Nellore, S.~S.~Pufu and F.~D.~Rocha,
Phys. Rev. Lett. \textbf{101} (2008), 131601
doi:10.1103/PhysRevLett.101.131601
[arXiv:0804.1950 [hep-th]].

\bibitem{Gursoy:2008bu}
U.~Gursoy, E.~Kiritsis, L.~Mazzanti and F.~Nitti,
Phys. Rev. Lett. \textbf{101} (2008), 181601
doi:10.1103/PhysRevLett.101.181601
[arXiv:0804.0899 [hep-th]].

\bibitem{Gursoy:2008za}
U.~Gursoy, E.~Kiritsis, L.~Mazzanti and F.~Nitti,
JHEP \textbf{05} (2009), 033
doi:10.1088/1126-6708/2009/05/033
[arXiv:0812.0792 [hep-th]].

\bibitem{Gursoy:2009jd}
U.~Gursoy, E.~Kiritsis, L.~Mazzanti and F.~Nitti,
Nucl. Phys. B \textbf{820} (2009), 148-177
doi:10.1016/j.nuclphysb.2009.05.017
[arXiv:0903.2859 [hep-th]].

\bibitem{Gursoy:2009kk}
U.~Gursoy, E.~Kiritsis, G.~Michalogiorgakis and F.~Nitti,
JHEP \textbf{12} (2009), 056
doi:10.1088/1126-6708/2009/12/056
[arXiv:0906.1890 [hep-ph]].

\bibitem{Noronha:2009ud}
J.~Noronha,
Phys. Rev. D \textbf{81} (2010), 045011
doi:10.1103/PhysRevD.81.045011
[arXiv:0910.1261 [hep-th]].

\bibitem{DeWolfe:2010he}
O.~DeWolfe, S.~S.~Gubser and C.~Rosen,
Phys. Rev. D \textbf{83} (2011), 086005
doi:10.1103/PhysRevD.83.086005
[arXiv:1012.1864 [hep-th]].

\bibitem{DeWolfe:2011ts}
O.~DeWolfe, S.~S.~Gubser and C.~Rosen,
Phys. Rev. D \textbf{84} (2011), 126014
doi:10.1103/PhysRevD.84.126014
[arXiv:1108.2029 [hep-th]].

\bibitem{Cai:2012xh}
R.~G.~Cai, S.~He and D.~Li,
JHEP \textbf{03} (2012), 033
doi:10.1007/JHEP03(2012)033
[arXiv:1201.0820 [hep-th]].



\bibitem{Yang:2014bqa}
Y.~Yang and P.~H.~Yuan,
JHEP \textbf{11} (2014), 149
doi:10.1007/JHEP11(2014)149
[arXiv:1406.1865 [hep-th]].


\bibitem{Finazzo:2016psx}
S.~I.~Finazzo, R.~Rougemont, M.~Zaniboni, R.~Critelli and J.~Noronha,
JHEP \textbf{01} (2017), 137
doi:10.1007/JHEP01(2017)137
[arXiv:1610.01519 [hep-th]].

\bibitem{Knaute:2017opk}
J.~Knaute, R.~Yaresko and B.~K\"ampfer,
Phys. Lett. B \textbf{778} (2018), 419-425
doi:10.1016/j.physletb.2018.01.053
[arXiv:1702.06731 [hep-ph]].

\bibitem{Sin:2007ze}
S.~J.~Sin,
JHEP \textbf{10} (2007), 078
doi:10.1088/1126-6708/2007/10/078
[arXiv:0707.2719 [hep-th]].

\bibitem{Colangelo:2010pe}
P.~Colangelo, F.~Giannuzzi and S.~Nicotri,
Phys. Rev. D \textbf{83} (2011), 035015
doi:10.1103/PhysRevD.83.035015
[arXiv:1008.3116 [hep-ph]].

\bibitem{Ballon-Bayona:2020xls}
A.~Ballon-Bayona, H.~Boschi-Filho, E.~Folco Capossoli and D.~M.~Rodrigues,
[arXiv:2006.08810 [hep-th]].

\bibitem{He:2013qq}
S.~He, S.~Y.~Wu, Y.~Yang and P.~H.~Yuan,
JHEP \textbf{04} (2013), 093
doi:10.1007/JHEP04(2013)093
[arXiv:1301.0385 [hep-th]].

\bibitem{Yang:2015aia}
Y.~Yang and P.~H.~Yuan,
JHEP \textbf{12} (2015), 161
doi:10.1007/JHEP12(2015)161
[arXiv:1506.05930 [hep-th]].

\bibitem{Mahapatra:2020wym}
S.~Mahapatra, S.~Priyadarshinee, G.~N.~Reddy and B.~Shukla,
Phys. Rev. D \textbf{102} (2020) no.2, 024042
doi:10.1103/PhysRevD.102.024042
[arXiv:2004.00921 [hep-th]].

\bibitem{Dudal:2017max}
D.~Dudal and S.~Mahapatra,
Phys. Rev. D \textbf{96} (2017) no.12, 126010
doi:10.1103/PhysRevD.96.126010
[arXiv:1708.06995 [hep-th]].


\bibitem{Bohra:2019ebj}
H.~Bohra, D.~Dudal, A.~Hajilou and S.~Mahapatra,
Phys. Lett. B \textbf{801} (2020), 135184
doi:10.1016/j.physletb.2019.135184
[arXiv:1907.01852 [hep-th]].



\bibitem{Arefeva:2020uec}
I.~Y.~Aref'eva, A.~Patrushev and P.~Slepov,
JHEP \textbf{07} (2020), 043
doi:10.1007/JHEP07(2020)043
[arXiv:2003.05847 [hep-th]].

\bibitem{Li:2017tdz}
M.~W.~Li, Y.~Yang and P.~H.~Yuan,
Phys. Rev. D \textbf{96} (2017) no.6, 066013
doi:10.1103/PhysRevD.96.066013
[arXiv:1703.09184 [hep-th]].


\bibitem{Chen:2019rez}
X.~Chen, D.~Li, D.~Hou and M.~Huang,
JHEP \textbf{03} (2020), 073
doi:10.1007/JHEP03(2020)073
[arXiv:1908.02000 [hep-ph]].

\bibitem{Chen:2020ath}
X.~Chen, L.~Zhang, D.~Li, D.~Hou and M.~Huang,
[arXiv:2010.14478 [hep-ph]].

\bibitem{He:2020fdi}
S.~He, Y.~Yang and P.~H.~Yuan,
[arXiv:2004.01965 [hep-th]].

\bibitem{Mamani:2020pks}
L.~A.~H.~Mamani, C.~V.~Flores and V.~T.~Zanchin,
Phys. Rev. D \textbf{102} (2020) no.6, 066006
doi:10.1103/PhysRevD.102.066006
[arXiv:2006.09401 [hep-th]].

\bibitem{Arefeva:2020vae}
I.~Y.~Aref'eva, K.~Rannu and P.~Slepov,
JHEP \textbf{07}, 161 (2021)
doi:10.1007/JHEP07(2021)161
[arXiv:2011.07023 [hep-th]].

\bibitem{Arefeva:2020byn}
I.~Y.~Aref'eva, K.~Rannu and P.~Slepov,
JHEP \textbf{06}, 090 (2021)
doi:10.1007/JHEP06(2021)090
[arXiv:2009.05562 [hep-th]].

\bibitem{Bohra:2020qom}
H.~Bohra, D.~Dudal, A.~Hajilou and S.~Mahapatra,
Phys. Rev. D \textbf{103}, no.8, 086021 (2021)
doi:10.1103/PhysRevD.103.086021
[arXiv:2010.04578 [hep-th]].

\bibitem{Chen:2021gop}
X.~Chen, L.~Zhang and D.~Hou,
[arXiv:2108.03840 [hep-ph]].

\bibitem{Zhao:2021ogc}
Y.~Q.~Zhao and D.~Hou,
[arXiv:2108.08479 [hep-ph]].

\bibitem{Zhou:2021nbp}
J.~Zhou and J.~Ping,
[arXiv:2101.08105 [hep-th]].

\bibitem{Zhang:2020zrv}
H.~X.~Zhang and B.~W.~Zhang,
Chin. Phys. C \textbf{45}, no.4, 044104 (2021)
doi:10.1088/1674-1137/abdf43
[arXiv:2007.13580 [hep-ph]].

\bibitem{Matsui:1986dk}
  T.~Matsui and H.~Satz,
  Phys.\ Lett.\ B {\bf 178}, 416 (1986).
  doi:10.1016/0370-2693(86)91404-8


\bibitem{Qin:2015srf}
G.~Y.~Qin and X.~N.~Wang,
Int. J. Mod. Phys. E \textbf{24}, no.11, 1530014 (2015)
doi:10.1142/S0218301315300143
[arXiv:1511.00790 [hep-ph]].

\bibitem{Matsuo:2006ws}
  T.~Matsuo, D.~Tomino and W.~Y.~Wen,
  JHEP {\bf 0610}, 055 (2006)
  doi:10.1088/1126-6708/2006/10/055
  [hep-th/0607178].

\bibitem{Caceres:2006dj}
  E.~Caceres and A.~Guijosa,
  JHEP {\bf 0611}, 077 (2006)
  doi:10.1088/1126-6708/2006/11/077
  [hep-th/0605235].

\bibitem{Rougemont:2015wca}
  R.~Rougemont, A.~Ficnar, S.~Finazzo and J.~Noronha,
  JHEP {\bf 1604}, 102 (2016)
  doi:10.1007/JHEP04(2016)102
  [arXiv:1507.06556 [hep-th]].

\bibitem{Cheng:2014fza}
  L.~Cheng, X.~H.~Ge and S.~Y.~Wu,
  Eur.\ Phys.\ J.\ C {\bf 76}, no. 5, 256 (2016)
  doi:10.1140/epjc/s10052-016-4096-7
  [arXiv:1412.8433 [hep-th]].

\bibitem{Mamo:2016xco}
  K.~A.~Mamo,
  Phys.\ Rev.\ D {\bf 94}, no. 4, 041901 (2016)
  doi:10.1103/PhysRevD.94.041901
  [arXiv:1606.01598 [hep-th]].

\bibitem{Zhang:2018mqt}
  Z.~q.~Zhang, K.~Ma and D.~f.~Hou,
  J.\ Phys.\ G {\bf 45}, no. 2, 025003 (2018)
  doi:10.1088/1361-6471/aaa097
  [arXiv:1802.01912 [hep-th]].

\bibitem{Arefeva:2020bjk}
I.~Y.~Aref'eva, K.~Rannu and P.~Slepov,
[arXiv:2012.05758 [hep-th]].


\bibitem{Akhavan:2008ep}
  A.~Akhavan, M.~Alishahiha, A.~Davody and A.~Vahedi,
  JHEP {\bf 0903}, 053 (2009)
  doi:10.1088/1126-6708/2009/03/053
  [arXiv:0811.3067 [hep-th]].

\bibitem{Hartnoll:2009ns}
  S.~A.~Hartnoll, J.~Polchinski, E.~Silverstein and D.~Tong,
  JHEP {\bf 1004}, 120 (2010)
  doi:10.1007/JHEP04(2010)120
  [arXiv:0912.1061 [hep-th]].

\bibitem{Giataganas:2013hwa}
  D.~Giataganas and H.~Soltanpanahi,
  Phys.\ Rev.\ D {\bf 89}, no. 2, 026011 (2014)
  doi:10.1103/PhysRevD.89.026011
  [arXiv:1310.6725 [hep-th]].

\bibitem{Sadeghi:2014lha}
  J.~Sadeghi and F.~Pourasadollah,
  Adv.\ High Energy Phys.\  {\bf 2014}, 670598 (2014)
  doi:10.1155/2014/670598
  [arXiv:1403.2192 [hep-th]].

\bibitem{Alishahiha:2012cm}
  M.~Alishahiha and H.~Yavartanoo,
  JHEP {\bf 1211}, 034 (2012)
  doi:10.1007/JHEP11(2012)034
  [arXiv:1208.6197 [hep-th]].

\bibitem{Kiritsis:2012ta}
  E.~Kiritsis,
  JHEP {\bf 1301}, 030 (2013)
  doi:10.1007/JHEP01(2013)030
  [arXiv:1207.2325 [hep-th]].

\bibitem{Kioumarsipour:2018tmf}
  M.~Kioumarsipour and J.~Sadeghi,
  J.\ Phys.\ G {\bf 45}, no. 8, 085001 (2018).
  doi:10.1088/1361-6471/aaca0f

\bibitem{Fadafan:2008bq}
  K.~Bitaghsir Fadafan, H.~Liu, K.~Rajagopal and U.~A.~Wiedemann,
  Eur.\ Phys.\ J.\ C {\bf 61}, 553 (2009)
  doi:10.1140/epjc/s10052-009-0885-6
  [arXiv:0809.2869 [hep-ph]].

\bibitem{Athanasiou:2010pv}
  C.~Athanasiou, P.~M.~Chesler, H.~Liu, D.~Nickel and K.~Rajagopal,
  Phys.\ Rev.\ D {\bf 81}, 126001 (2010)
  Erratum: [Phys.\ Rev.\ D {\bf 84}, 069901 (2011)]
  doi:10.1103/PhysRevD.81.126001, 10.1103/PhysRevD.84.069901
  [arXiv:1001.3880 [hep-th]].

\bibitem{AliAkbari:2011ue}
  M.~Ali-Akbari and U.~Gursoy,
  JHEP {\bf 1201}, 105 (2012)
  doi:10.1007/JHEP01(2012)105
  [arXiv:1110.5881 [hep-th]].

\bibitem{Fadafan:2012qu}
  K.~B.~Fadafan and H.~Soltanpanahi,
  JHEP {\bf 1210}, 085 (2012)
  doi:10.1007/JHEP10(2012)085
  [arXiv:1206.2271 [hep-th]].

\bibitem{Atashi:2016cso}
  M.~Atashi, K.~Bitaghsir Fadafan and M.~Farahbodnia,
  Eur.\ Phys.\ J.\ C {\bf 77}, no. 3, 175 (2017)
  doi:10.1140/epjc/s10052-017-4742-8
  [arXiv:1606.09491 [hep-th]].

\bibitem{Hou:2021own}
D.~Hou, M.~Atashi, K.~Bitaghsir Fadafan and Z.~q.~Zhang,
Phys. Lett. B \textbf{817} (2021), 136279
doi:10.1016/j.physletb.2021.136279

\bibitem{NataAtmaja:2010hd}
  A.~Nata Atmaja and K.~Schalm,
  JHEP {\bf 1104}, 070 (2011)
  doi:10.1007/JHEP04(2011)070
  [arXiv:1012.3800 [hep-th]].

\bibitem{Arefeva:2020jvo}
I.~Y.~Aref'eva, A.~A.~Golubtsova and E.~Gourgoulhon,
JHEP \textbf{04} (2021), 169
doi:10.1007/JHEP04(2021)169
[arXiv:2004.12984 [hep-th]].

\bibitem{Gubser:2006nz}
  S.~S.~Gubser,
  Nucl.\ Phys.\ B {\bf 790}, 175 (2008)
  doi:10.1016/j.nuclphysb.2007.09.017
  [hep-th/0612143].

\bibitem{Nakano:2006js}
  E.~Nakano, S.~Teraguchi and W.~Y.~Wen,
  Phys.\ Rev.\ D {\bf 75}, 085016 (2007)
  doi:10.1103/PhysRevD.75.085016
  [hep-ph/0608274].

\bibitem{Talavera:2006tj}
  P.~Talavera,
  JHEP {\bf 0701}, 086 (2007)
  doi:10.1088/1126-6708/2007/01/086
  [hep-th/0610179].


\bibitem{Roy:2009sw}
  S.~Roy,
  Phys.\ Lett.\ B {\bf 682}, 93 (2009)
  doi:10.1016/j.physletb.2009.10.095
  [arXiv:0907.0333 [hep-th]].

\bibitem{Panigrahi:2010cm}
  K.~L.~Panigrahi and S.~Roy,
  JHEP {\bf 1004}, 003 (2010)
  doi:10.1007/JHEP04(2010)003
  [arXiv:1001.2904 [hep-th]].

\bibitem{Chernicoff:2012iq}
  M.~Chernicoff, D.~Fernandez, D.~Mateos and D.~Trancanelli,
  JHEP {\bf 1208}, 100 (2012)
  doi:10.1007/JHEP08(2012)100
  [arXiv:1202.3696 [hep-th]].

\bibitem{Chakraborty:2014kfa}
  S.~Chakraborty and N.~Haque,
  JHEP {\bf 1412}, 175 (2014)
  doi:10.1007/JHEP12(2014)175
  [arXiv:1410.7040 [hep-th]].

\bibitem{Zhang:2019cxu}
Z.~q.~Zhang and X.~Zhu,
Eur. Phys. J. C \textbf{79} (2019) no.2, 107
doi:10.1140/epjc/s10052-019-6579-9

\bibitem{Andreev:2017bvr}
O.~Andreev,
Mod. Phys. Lett. A \textbf{33}, no.06, 1850041 (2018)
doi:10.1142/S0217732318500414
[arXiv:1707.05045 [hep-ph]].

\bibitem{Bellwied:2015rza}
R.~Bellwied, S.~Borsanyi, Z.~Fodor, J.~G\"unther, S.~D.~Katz, C.~Ratti and K.~K.~Szabo,
Phys. Lett. B \textbf{751} (2015), 559-564
doi:10.1016/j.physletb.2015.11.011
[arXiv:1507.07510 [hep-lat]].

\bibitem{Borsanyi:2012cr}
S.~Borsanyi, G.~Endrodi, Z.~Fodor, S.~D.~Katz, S.~Krieg, C.~Ratti and K.~K.~Szabo,
JHEP \textbf{08} (2012), 053
doi:10.1007/JHEP08(2012)053
[arXiv:1204.6710 [hep-lat]].


\bibitem{Breitenlohner:1982jf}
P.~Breitenlohner and D.~Z.~Freedman,
Annals Phys. \textbf{144} (1982), 249
doi:10.1016/0003-4916(82)90116-6

\bibitem{Gubser:2000nd}
S.~S.~Gubser,
Adv. Theor. Math. Phys. \textbf{4} (2000), 679-745
doi:10.4310/ATMP.2000.v4.n3.a6
[arXiv:hep-th/0002160 [hep-th]].


\end{thebibliography}
\end{document}